\documentclass[12pt]{article}
\usepackage[left=2cm, right=2cm, bottom=2cm ,top = 2cm, footskip=1cm]{geometry}
\usepackage[font=small,labelfont=bf]{caption}

\usepackage{setspace}
\let\counterwithin\relax
\usepackage{chngcntr}

\usepackage{hyperref}
\usepackage[numbers,sort&compress]{natbib}
\setcitestyle{open={(},close={)}}

\usepackage{graphicx}
\usepackage{amsmath}

\usepackage{multirow} % for making cells that span more than 1 row in a column

\usepackage{caption} 
\usepackage{subcaption}
\usepackage{tabularx}

\usepackage[section]{placeins}

\title{More variable circadian rhythms in epilepsy captured by long-term heart rate recordings from wearable sensors}

\author{Billy C. Smith$^{1*}$, Christopher Thornton$^{1, 2}$, Rachel E. Stirling$^{5}$,\\Guillermo M. Besn\'e$^{1}$, Sarah J. Gascoigne$^{1}$, Nathan Evans$^{1}$, Peter N. Taylor$^{1,3,4}$,\\ Karoline Leiberg$^{1}$, Philippa J. Karoly$^{5}$, Yujiang Wang$^{1,3,4*}$}

\begin{document}

\maketitle

% author affiliations
\begin{enumerate}
\item{CNNP Lab (www.cnnp-lab.com), School of Computing, Newcastle University, Newcastle upon Tyne, United Kingdom, NE4 5TG}
\item{School of Computing, Engineering \& Digital Technologies, Teesside University, Middlesbrough, Tees Valley, TS1 3BX }
\item{Faculty of Medical Sciences, Newcastle University, Newcastle upon Tyne, United Kingdom, NE2 4HH}
\item{UCL Queen Square Institute of Neurology, Queen Square, London, United Kingdom, WC1N 3BG}
\item{Graeme Clark Institute and Department of Biomedical Engineering, The University of Melbourne, Victoria 3010, Australia}
\end{enumerate}

\begin{center}
* Email: b.smith16@ncl.ac.uk or Yujiang.Wang@ncl.ac.uk%, Tel: 0191 208 7972
\end{center}

%\noindent We confirm that we have read the Journal’s position on issues involved in ethical publication and affirm that this report is consistent with those guidelines.

%\noindent\textbf{Keywords:} intra-individual, variability, day-to-day variability, seizure, wearables, heart rate, circadian, rhythm, circadian disruption

%\noindent\textbf{Classification:} Biological Sciences, Physiology

%\noindent This article is currently available as a pre-print on arXiv (arXiv:2411.04634) under an arxiv.org perpetual, non-exclusive license. 

\newpage

\section*{Abstract}
% The circadian rhythm aligns life on earth with the 24-hour light-dark cycle, but can be disrupted by modern lifestyles. While many diseases, including epilepsy, are modulated by circadian rhythms, it is unclear if these conditions involve inherent disruptions to the circadian rhythm, or if such disruptions are behavioural adaptations to disease symptoms. Using wearable smartwatch heart rate data (Fitbit) from 143 people with epilepsy (PWE) and 31 controls retrospectively, we extract circadian rhythms from their heart rate and estimate intra-individual variability of circadian period, acrophase and amplitude. We found that PWE have more variable and less stable heart rate circadian rhythms than controls ($z\approx3, p<=0.003$), however this was unrelated to seizure frequency or occurrence. These results suggest that circadian disruptions may stem from underlying epilepsy pathology rather than being a response to symptoms, highlighting their potential broader role in health conditions.

\textbf{Objective}: The circadian rhythm synchronizes physiological and behavioural patterns with the 24-hour light-dark cycle. Disruption to the circadian rhythm is linked to various health conditions, though optimal methods to describe these disruptions remain unclear. An emerging approach is to examine the intra-individual variability in measurable properties of the circadian rhythm over extended periods. Epileptic seizures are modulated by circadian rhythms, but the relevance of circadian rhythm disruption in epilepsy remains unexplored. Our study investigates intra-individual circadian variability in epilepsy and its relationship with seizures.

\textbf{Methods}: We retrospectively analyzed over 70,000 hours of wearable smartwatch data (Fitbit) from 143 people with epilepsy (PWE) and 31 healthy controls. Circadian oscillations in heart rate time series were extracted, daily estimates of circadian period, acrophase, and amplitude properties were produced, and estimates of the intra-individual variability of these properties over an entire recording were calculated.

\textbf{Results}: PWE exhibited greater intra-individual variability in period (76~min vs. 57 min, $d=0.66, p<0.001$) and acrophase (64~min vs. 48~min, $d=0.49, p=0.004$) compared to controls, but not in amplitude (2~bpm, $d=-0.15, p=0.49$). Variability in circadian properties showed no correlation with seizure frequency, nor any differences between weeks with and without seizures.

\textbf{Significance}: For the first time, we show that heart rate circadian rhythms are more variable in PWE, detectable via consumer wearable devices. However, no association with seizure frequency or occurrence was found, suggesting that this variability might be underpinned by the epilepsy aetiology rather than being a seizure-driven effect.

%\section*{Significance}
%The circadian rhythm aligns life on earth with the 24-hour light-dark cycle, but can be disrupted by modern lifestyles. While many diseases, including epilepsy, are modulated by circadian rhythms, it is unclear if these conditions involve inherent disruptions to the circadian rhythm, or if such disruptions are behavioural adaptations to disease symptoms. This study found that people with epilepsy have more variable and less stable heart rate circadian rhythms than controls, regardless of seizure frequency or occurrence. These results suggest that circadian disruptions may stem from underlying disease pathology rather than being a response to symptoms, highlighting their potential broader role in health conditions.

\section*{Key points} %up to 140 characters each 
\begin{itemize}
    
    %\item Modern lifestyles, and some diseases, are known to disrupt the circadian rhythm. It is unclear whether epilepsy is such a condition. 
    
    %\item We measured circadian rhythm disruption using continuous wearable heart rate recordings ranging from months to years in PWE and controls.

    \item Relative to controls, the circadian rhythm of heart rate was more variable for PWE, measured using commercial wearable devices. 
    
    \item This provides evidence for long-term and persistent circadian disruption in epilepsy.
    
    %\item Relative to controls, we found a more variable rhythm for PWE, though this was not related to seizure occurrence or frequency.

    \item However, increased intra-individual circadian variability was not related to seizure occurrence or frequency.
    
     %\item Future studies should investigate whether increased variability is instead associated with the underlying disease aetiology, or with other factors such as ASMs, sleep, or comorbidities.

     \item The underlying aetiology may instead be associated, or factors such as ASMs or comorbidities may play an important role.
    
\end{itemize}

\newpage

\section{Introduction}

The circadian rhythm aligns our physiology and behaviour to the 24-hour environmental light-dark cycle. A stable circadian rhythm is thought to be important for overall health, and disruptions to the circadian rhythm have been associated with various conditions, including sleep disorders \citep{Burgess2017, Fishbein2021}, psychiatric disorders \citep{Ali2023, Carr2018, Song2024, Burns2024}, and neurological disorders \citep{Logan2019}. However, it remains unclear how circadian disruption should be assessed over the long term - many methods are available depending upon context \citep{Vetter2018, AsgariTarghi2019, Baron2014}.

One promising and easy-to-interpret approach is to assess the intra-individual variability of a set of fundamental descriptive properties of the circadian rhythm, for example amplitude, measured across multiple consecutive days. This approach has also been referred to as `day-to-day variability' \citep{Fossion2017, GarciaIglesias2023} or `circadian variability' \cite{Carr2018}. Intra-individual variability is designed to detect persistent irregularities in circadian rhythms over long-term physiological recordings. Such data are most easily obtained from peripheral measures that exhibit circadian oscillations using wearable devices (such as heart rate, physical activity, and skin temperature), making this approach a cost-effective and scalable methodology to study circadian disruption, complementary with existing methods.

Epilepsy is a neurological disorder characterized by recurrent seizures \citep{Devinsky2018}, which are closely associated with the circadian rhythm \citep{Karoly2021, Bazhanova2022, Hofstra2009, Thornton2024a, Gascoigne2023, Stirling2023, Najar2024}. Sleep patterns are inherently tied to the circadian rhythm, and the relationship between sleep and seizures has been reported upon widely \citep{Degen1980, Bazil2003, Derry2013, Grayson2018, DellAquila2022}. However, the role of persistent, long-term circadian disruption in epilepsy remains unstudied. Consequently, we propose the assessment of intra-individual circadian variability in wearable heart rate recordings sourced from people with epilepsy (PWE), which could provide valuable insights for clinical applications, such as enhancing wearable seizure prediction \citep{Vieluf2022} and guiding chronotherapeutic treatments \citep{Bazhanova2022, Najar2024}, including ASM alert systems. Additionally, this novel application of intra-individual variability as a measure of circadian disruption in epilepsy could support its broader use as a wearable clinical biomarker.

Here, we measure the intra-individual variability of three descriptive properties of the circadian rhythm of heart rate using long-term wearable-derived heart rate recordings from 143 PWE and 31 controls. We seek to establish whether circadian variability is greater in epilepsy and, if so, whether a relationship exists between variability and seizure frequency and occurrence.

\section{Materials \& methods}

\subsection{Participant data}

Wearable smartwatch data from people with epilepsy (PWE) and controls were used retrospectively, sourced from the observational ``Tracking Seizure Cycles" study \citep{Karoly2021}.  Adults with epilepsy were recruited to this study by referral from collaborating epilepsy specialists in tertiary referral epilepsy clinics. Inclusion criteria were diagnosis of epilepsy from a specialised epileptologist, uncontrolled/partially controlled seizures as determined by their neurologist, and that they were deemed capable of keeping a reliable seizure diary. Additionally, \cite{Karoly2021} recruited control participants without epilepsy from their colleagues, friends and relatives, with no randomisation or blinding performed. Their study was approved by the St Vincent’s Hospital Human Research Ethics Committee (HREC 009.19). All participants provided written informed consent.

For both PWE and controls, data were collected via a wearable smartwatch (Fitbit), continuously measuring heart rate via photoplethysmography (PPG) at 5~s resolution. PWE participating also used a mobile device to manually report clinically-apparent seizures using the freely-available Seer App for either the entirety, or a subset of, the study period.

Our retrospective analysis of this data was approved by the Newcastle University Ethics Committee (40679/2023).

\subsection{Measuring circadian disruption using physiological time-series: the `intra-individual variability' approach}\label{section:circ-var}

\begin{figure}
    \centering
    \includegraphics[scale=1]{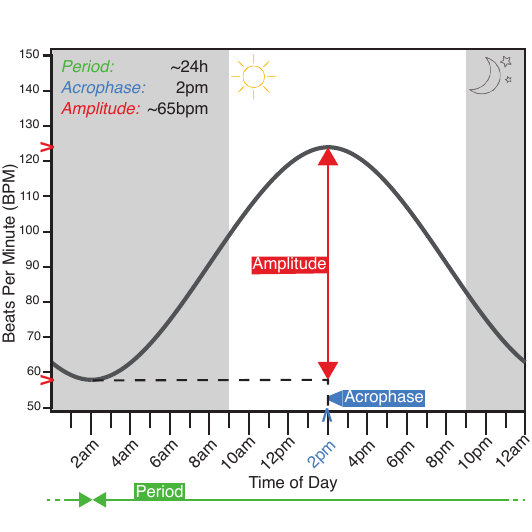}
    \caption{An illustrative schematic of one `cycle' of the circadian rhythm of heart rate (CRHR) with the 3 descriptive properties used in this study labelled. Period reflects the duration of one cycle of the modelled CRHR (approximately 24 hours). Acrophase is defined as the time-of-day at peak cycle magnitude, reflecting rhythm ‘timing’.
    Amplitude is a measure of circadian ‘strength’, defined as the difference in magnitude between cycle peak and previous trough. 
}
    \label{fig:methods1}
\end{figure}

Capturing the intra-individual variability of the circadian rhythm involves the daily measurement of three descriptive circadian properties, which are illustrated in Figure~\ref{fig:methods1}. For each participant, the circadian rhythm of heart rate (CRHR) was extracted from their wearable recording. Three circadian properties were then computed for each daily cycle in the CRHR. Finally, estimates of the average (statistical mean) and variability (statistical standard deviation) of each property over the recording were produced. Figure~\ref{fig:methods2} provides an overview of this process, and further details are provided below and in Supplementary~\ref{suppl:method_details}.

\subsubsection{Accounting for gaps in recording}
%Outliers in the heart rate timeseries with an absolute modified z-score ($z = 0.6745() / MAD$ greater than 3.5 were removed. 
To calculate the circadian properties of the heart rate, a continuous signal was required. Any gaps shorter than 8 hours were linearly interpolated, and the signal was split up into `runs' of data between remaining gaps (Figure~\ref{fig:methods2}B).
26 participants (21 PWE, 5 control) without any `runs' longer than 7 days were excluded from further analysis. A more detailed explanation and justification for this approach is provided in Supplementary~\ref{supp:runs_and_segmenting_extra_detail}. 

\subsubsection{Extracting the circadian rhythm}

For each `run' of data, singular spectrum analysis (SSA) was applied to decompose the continuous raw heart-rate time series into components of distinct frequency (Figure~\ref{fig:methods2}C-E(i)). 
%
%SSA was found to be the best-performing algorithm from a set of similar methods (see Supplementary~\ref{suppl:modelling_methods_comparison}).
SSA was found to perform reliably in comparison to similar algorithms (see Supplementary~\ref{suppl:modelling_methods_comparison}).
To determine which component represented the CRHR, Fourier analysis was performed on each component to determine its central frequency. The component with period (inverse of frequency) closest to 24 hours was selected as the circadian component (Figure \ref{fig:methods2}E(i)). \\

\subsubsection{Computing circadian properties}

The extracted CRHR was then split-up into the individual (approximately daily) circadian cycles (Figure \ref{fig:methods2}E), and the period, acrophase and amplitude were computed for each cycle (Figure \ref{fig:methods2}G). To achieve this, the Hilbert transform was applied to the extracted CRHR, producing a complex-valued analytic signal, from which the circadian ``phase series'' (Supplementary~\ref{suppl:hilbert}), a measure of circadian cycle progression, was derived (Figure \ref{fig:methods2}E(ii)). The rhythm was then split into daily cycles at each trough in the phase-series. For each individual cycle, \textit{period} was computed as the duration between the first and last time point of the cycle, \textit{acrophase} as the time of day at zero cycle phase, and \textit{amplitude} as the difference in magnitude between acrophase and trough (Figure~\ref{fig:methods1}). \\

\subsubsection{Calculating the intra-individual variability of circadian properties}\label{section:variability}

\begin{figure}
    \centering
    % TODO SSA, per-cycle values, segmenting? figure
    \includegraphics[scale=0.75]{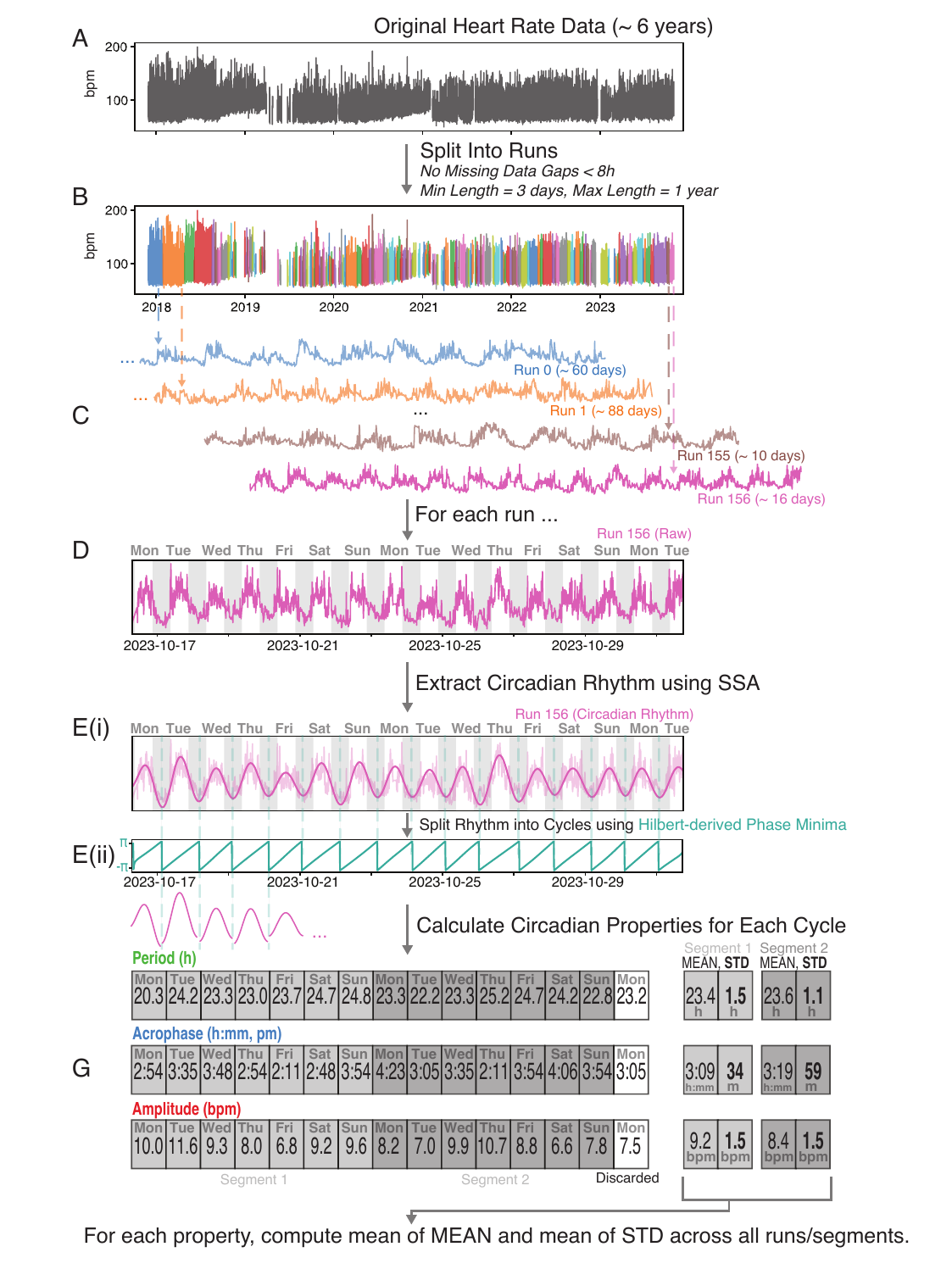}
    \caption{
    Participant heart rate data (A) was split into runs between $>$8h gaps of missing data (B). 
    SSA extracted the CRHR from each run (C to E(i)). 
    Troughs in the circadian phase (E(ii)) were used as reference points for splitting the extracted CRHR into individual cycles.
    Circadian properties were calculated for each cycle and grouped into segments of seven days (G).   
    The mean and standard deviation of each property was then calculated within each 7-day segment. 
    The mean across all segments in a subject produces the final intra-individual average and variability values. 
    For example, if this example participant had only these two segments, their intra-individual variability of acrophase  would be $(34 + 59)/2 = 46.5$ minutes. 
    %
    %These are shown in Figures~\ref{fig:results1}-\ref{fig:results2}.
    }
    \label{fig:methods2}
\end{figure}

Circadian properties for each cycle were grouped into consecutive, non-overlapping segments of seven days. The standard deviation was calculated using all seven days in each segment and property (Figure~\ref{fig:methods2}G), and the summary \textit{intra-individual variability} was calculated from the average of the standard deviations across all segments. The \textit{intra-individual average} of circadian properties was calculated in the same way from the average of the means across segments. 
The fixed segment size of seven days prevents any bias introduced by varying recording duration and `run' length between participants when calculating the mean and standard deviation (see Supplementary \ref{supp:runs_and_segmenting_extra_detail} for more detail).

\subsection{Statistical analysis}\label{stats}
We calculate p-values for reference only, and all reported values are raw, uncorrected by FDR. 
A non-parametric two-sided Wilcoxon rank-sum test was used to compare the intra-individual average and variability of each circadian property between PWE and controls. A non-parametric test was chosen to account for the imbalance between PWE and control sample sizes, but to further control for this a random sub-sampling test was performed over 10,000 iterations, where the whole control sample was compared to a random sub-sample of 28 PWE using the same statistical method as above. Sampling was without replacement within an iteration, but with replacement across iterations. %To ensure that the longer overall recording duration in PWE relative to controls was not impacting our results, we re-ran the analysis with every subjects sub-sampled to the same number of segments. The observations remain the same, see Supplementary~\ref{suppl:methods_details}.

Pearson’s correlation was used to test whether an individual's seizure frequency was correlated with their intra-individual variability calculated over segments occurring between the start and end of their seizure diary. Seizure frequency was defined in terms of the average number of seizures per week of recording data post-segmenting. The distribution of seizure frequency was highly skewed, so a log10 transformation was applied for the correlation analysis. 

To test whether having one or more seizures in a 7-day period was associated with greater intra-individual variability, each 7-day segment occurring within a participant's seizure diary was classified as either seizure-containing or seizure-free according to the seizure diary. For PWE with at least 5 seizure-free segments and at least 5 seizure-containing segments, intra-individual circadian average and variability values were calculated over seizure-free and seizure-containing segments separately. These values were then compared using a two-sided Wilcoxon signed-rank paired test. 
  
Effect sizes were reported using Cohen's d throughout. 

\clearpage
\section{Results}

Before processing, data were available for n=164 people with epilepsy (PWE) and n=36 controls. After processing, 143 PWE and 31 controls remained for analysis. We analysed and compared intra-individual variability of circadian rhythm properties (see Methods) calculated using wearable heart rate data (FitBit), between PWE and controls. 

Briefly, we isolated the circadian rhythm component from the raw heart rate recordings, and measured a range of properties (period, acrophase, amplitude) for each cycle. Intra-individual variability in circadian properties were then obtained across segments of successive cycles. An increased intra-individual variability in any property can be interpreted as a less ``stable'' circadian rhythm for the individual.

Supplementary table~\ref{tab:melbourne_data_after_processing} lists the demographics and other metadata of this cohort. An average of 163~days (sd = 183) were discarded for each subject, resulting in a median recording duration of 215.8~days (IQR = 580.9) for PWE and 125.1~days (IQR = 260.5) for controls. Although recording duration was shorter in controls compared to PWE, we found no relationship between (post-segmenting) recording duration and intra-individual variability in any circadian property (Supplementary~\ref{suppl:rec_duration}). There were no age ($t=-0.215, p=0.84$) or sex ($\chi^2=2.777, p=0.427$) differences between PWE and controls, and there was no association between intra-individual variability and age, sex or epilepsy sub-type (see Supplementary~\ref{suppl:age_sex}).  

%It should be noted that `day' refers to a unit of 24 hours that were not necessarily consecutive, i.e a single discarded `day' of recording could refer to four temporally isolated six-hour intervals of recording. \\

\subsection{The circadian rhythm of heart rate is more variable for people with epilepsy}

\begin{figure}
    \centering
    \includegraphics[scale=1]{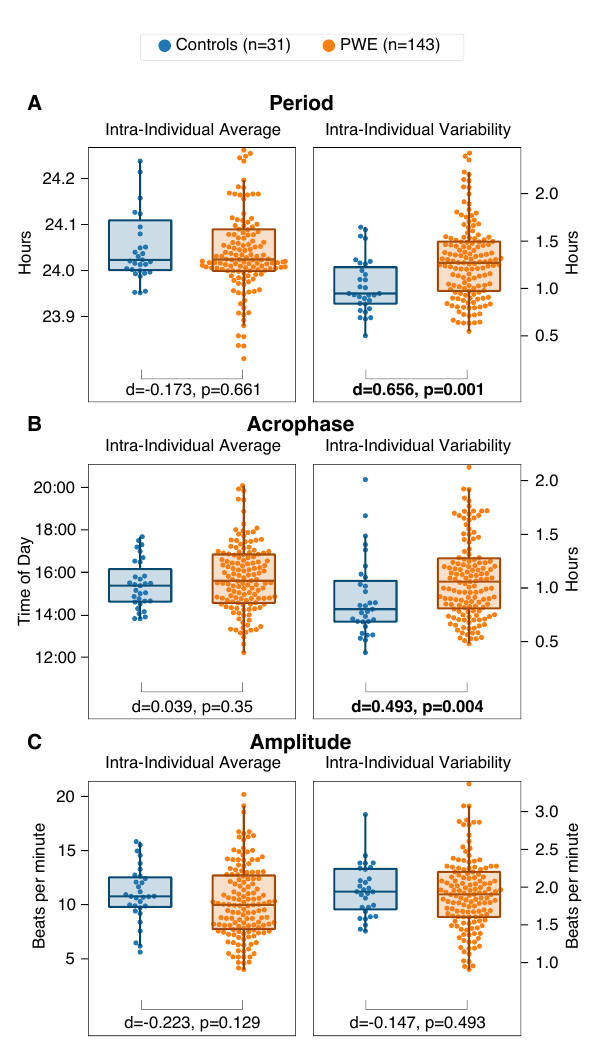}
    \caption{\textbf{Comparison of the distribution of the intra-individual average and variability of circadian properties between PWE and controls.} 
    Period, acrophase and amplitude are defined in Figure~\ref{fig:methods1}.
    Effect sizes are measured in Cohen's d and p-values are based on the two-sided Wilcoxon rank-sum test for comparison between PWE and Controls.
     Each subpanel has been zoomed-in to omit some outliers; outliers were not excluded for statistical testing.
    }
    
    \label{fig:results1}
\end{figure}

The intra-individual variability in period (76 min \textit{vs.} 57 min, $d=0.66, p<0.001$, Figure~\ref{fig:results1}A) and acrophase (64 min \textit{vs.} 48 min, $d=0.49, p=0.004$, Figure~\ref{fig:results1}B) was increased for PWE compared to controls. However, there was no difference in intra-individual variability of amplitude between PWE and controls ($\sim$2 bpm, $d=-0.15, p=0.49$, Figure~\ref{fig:results1}C).

For reference, we also present the results for intra-individual average in Figure~\ref{fig:results1}, but found no substantial difference between PWE and controls (period: $d=-0.17, p=0.66$; acrophase: $d=0.04, p=0.35$; amplitude: $d=-0.22, p=0.13$). 

All results held after applying a repeated sub-sampling test with randomly selected samples of PWE equal in size to controls (see Supplementary~\ref{suppl:subsampling}).

\subsection{Intra-individual variability does not correlate with increased seizure frequency}

\begin{figure}
    
    \centering
    \includegraphics[scale=1]{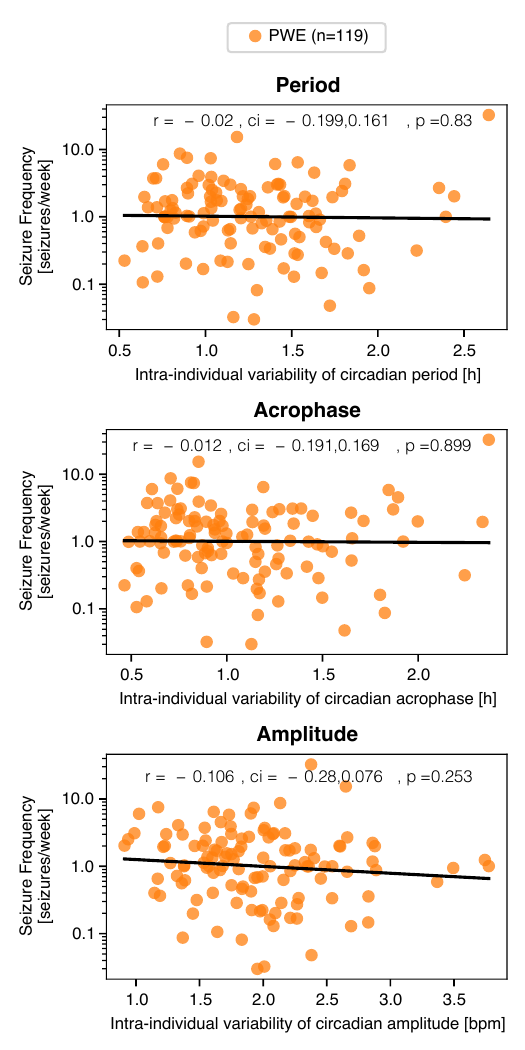}
    \caption{
        \textbf{Scatter plots between intra-individual variability of circadian properties and seizure frequency.}
        Seizure frequency is in units of seizures per week on log10 scale.}\label{fig:results2}
\end{figure}

Next, the correlation between seizure frequency and intra-individual variability was investigated (Figure~\ref{fig:results2}). PWE without any seizures recorded or with outlying intra-individual variability (absolute z-score $>$ 3) in any property were removed, leaving 119 PWE. We found that intra-individual variability of any property was not correlated with seizure frequency ($|r| \leq 0.1$).

\subsection{Intra-individual variability does not differ between weeks with and without seizures}

\begin{figure}
    \centering
    \includegraphics[scale=1]{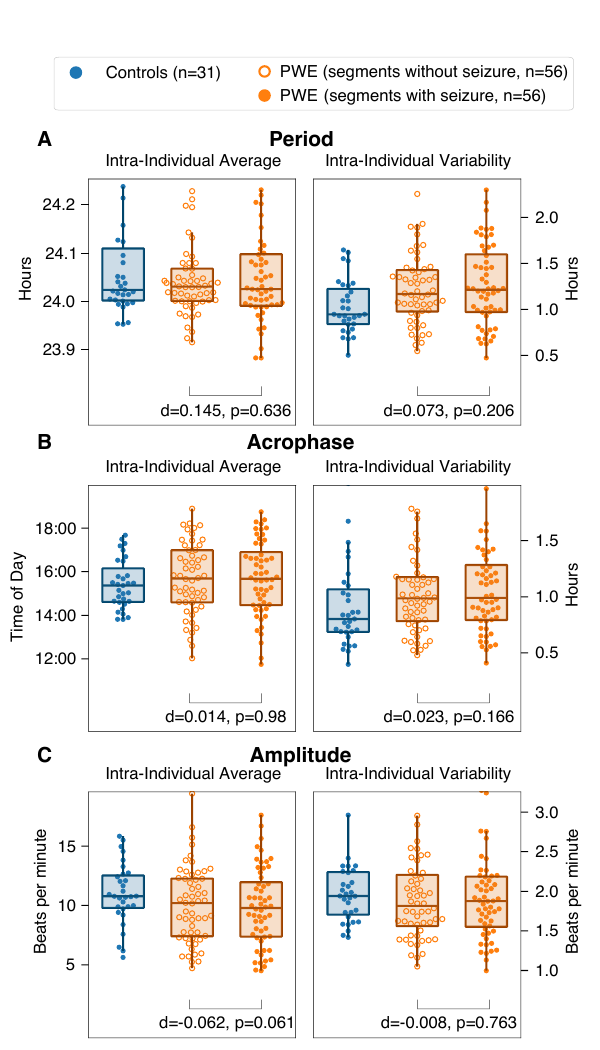}
     \caption{\textbf{Intra-individual average and variability of circadian properties calculated over segments with and without seizures.}
     Controls are shown for reference and are not included in the statistics shown.
     The two-sided Wilcoxon signed-rank paired test was used for comparison of the intra-individual average and variability calculated using segments with and without seizures.
     Each subpanel has been zoomed-in to omit outliers; outliers were not excluded prior to statistical testing.}\label{fig:results3}
    
\end{figure}

We investigated whether intra-individual variability differed between seizure-containing or seizure-free 7-day segments (Figure~\ref{fig:results3}). Individuals without at least 5 seizure-free segments and at least 5 seizure-containing segments were excluded, leaving 56 PWE. There was no paired difference in intra-individual variability or average of any circadian property between seizure-free \textit{vs.} seizure-containing segments.

\subsection{Weekend and time of year effects}

To supplement our findings, we investigated (Supplementary~\ref{suppl:weekly_annual}) whether weekdays and weekends, as well as the time of year were associated with the observed increased intra-individual variability in PWE.

We observed a difference between weekends and weekdays in almost all measures (intra-individual averages and variabilities) for both controls and PWE (Fig~\ref{fig:supp_weekend_within_group}). However, the previously observed increases in intra-individual variability in period and acrophase for PWE remained regardless of weekday or weekend (Fig.~\ref{fig:supp_weekend_between_groups}). This observation was additionally confirmed by a mixed effects model, where we tested for the interaction of weekday/weekend with control/PWE. More details and full statistical models can be found in Suppl.~\ref{suppl:weekly}.

We also tested for a seasonal effect in a similar manner, where we found a less pronounced effect overall. We again confirmed in that the observed increase in intra-individual variability on period and acrophase were not driven by seasonal effects or interactions. More details and full statistical models can be found in Suppl.~\ref{suppl:annual}.

In summary, whilst weekly and seasonal effects are present in intra-individual averages and variability, these do not drive our main results of increased intra-individual variability in PWE. 

\section{Discussion}

%% SUMMARY: Overview
This study analysed the intra-individual variability of three properties of the circadian rhythm using long-term wearable heart rate data from 143 PWE and 31 controls. We found increased variability in circadian period and acrophase for PWE, providing crucial evidence of circadian disruption in epilepsy. However, we found no evidence of an association between intra-individual variability and seizure frequency, or occurrence, at the population level.

%% SUMMARY: Importance for epilepsy
Circadian disruption has a bi-directional relationship with health \citep{Abbott2020} and is associated with negative health outcomes \citep{Baron2014} and mortality over the long term \citep{Zuurbier2015}. It is therefore likely to have compounding health implications for PWE, so the results presented here should motivate the development of interventions to reduce the co-morbidities associated with circadian disruption for PWE \citep{Latreille2018}. 

%% Potential links between epilepsy and circadan disruption
We found no evidence of any link between circadian disruption and seizures but consider a link with epilepsy aetiology possible. Epilepsy often has a genetic cause \citep{Devinsky2018} and many related genes have been identified. Associations between epilepsy and abnormal expression of the genes underlying the circadian rhythm have also been reported \citep{Khan2018, Kreitlow2022}. For example, evidence from human \textit{ex vivo} pathological tissue suggests a link between decreased expression of BMAL1 and temporal lobe epilepsy \citep{Wu2021} - similar findings have been reported for the CLOCK gene \cite{Li2017}. Functionally, the circadian amplitude of neural activity has been found to be reduced in pathological areas using intracranial EEG \cite{Thornton2024a}. %Genetic disruption to the circadian rhythm in the brain likely also has downstream effects on the expression of the circadian rhythm in heart rate, which may explain the findings in this study. 
It is possible that any circadian rhythm disruption with a genetic basis at the brain level could be reciprocated at the heart, either genetically or as a downstream effect of disruption to the central circadian rhythm - such effects may explain the disruption to the circadian rhythm of heart rate for people with epilepsy that we observe here.
Moreover, co-morbidities of epilepsy such as anxiety and depression \citep{Leidy1999, Keezer2016} are also associated with circadian disruption \citep{Walker2020}. Within the epilepsy cohort we found that intra-individual variability has a wide range and considerable overlap with the control distribution: future work should investigate the extent to which these co-morbidities can explain our findings.

Misalignment of the circadian rhythm with the environmental light/dark cycle, particularly towards the evening, is one form of circadian disruption that has been associated with psychiatric disorders \citep{Burns2024, Baron2014}. `Chronotype' - an individual's innate preference as a morning or evening type - relates to such misalignment, and has been investigated in epilepsy \citep{Choi2016, Khan2018, Najar2024}. Precisely reporting on chronotype requires specific methodology so is outside the scope of our study, though average acrophase most closely resembles chronotype. No differences in acrophase were observed between PWE and controls, or between focal and generalised sub-types in our data (Supplementary~\ref{fig:epilepsy_type}). 
%Related to chronotype, \cite{Choi2016} report increased social jetlag in epilepsy. To this end, we also tested whether circadian properties and intra-individual variability differed depending on day of the week or weekends (Supplementary~\ref{suppl:weekly_annual}). We observed similar weekend effects across PWE and controls, suggesting that the increased intra-individual variability in PWE was not driven by stronger weekend effects in PWE. 
Incidentally, we did observe some differences in circadian rhythms between weekends and weekdays (Figure~\ref{fig:supp_weekend_within_group}) for both PWE and controls. In particular the differences in \textit{intra-individual average} period and acrophase may be related to the `social jetlag' phenomenon, which refers to differences in sleep and behavioural patterns on weekdays compared to weekends due to the constraints of work, education or other commitments \citep{Vetter2018}. These changes in \textit{intra-individual averages} were different for PWE compared to controls, which may support previous findings that social jetlag is exacerbated for PWE \citep{Choi2016}.  However, most weekday \textit{vs.} weekend effects in \textit{intra-individual variability} were independent of the control \textit{vs.} PWE effects (Figure~\ref{fig:supp_weekend_between_groups}, Table~\ref{tbl:weekend_LR}). Specifically, the highlighted increased intra-individual variability in PWE was not driven by the weekday/weekend effect in our data. It should be noted that insights into social jetlag are limited without inclusion of sleep measures in this study. Future work could test, in a more formal way, the mechanistic relationship between social jetlag, chronotypes, sleep, and our proposed measures of circadian variability.

Interestingly, in this study we observed that the median circadian period was marginally longer than 24 hours (24.02h) for both controls and PWE. In constant laboratory conditions without time cues such as light/dark cycles, the circadian rhythm persists in humans, but with a period of up to 25 hours \citep{Yamanaka2020}. In a laboratory setting where light/dark cycles are fixed, \citep{Duffy2011} report a variable circadian period of temperature and melatonin rhythms between participants, above and below 24 hours, that was shorter (24.09h) on average for females than males (24.19h). Our study was undertaken under no such controlled conditions: circadian period would certainly be entrained by light and other environmental cues, which may explain why our reported average is closer to 24 hours. Additionally, we did not observe any difference in average circadian period between males and females (Supplementary~\ref{suppl:age_sex}). Shifts in average circadian period may be associated with modulatory effects of longer, multi-day rhythms of heart rate \citep{Karoly2021}, or associated with the annual variations of circadian properties that we also observed (Section~\ref{suppl:annual}). Future work could explore the dynamics of circadian period over-time in detail and whether re-entrainment differs between PWE and controls.  

%Another co-incidental finding is that 
%Similarly, some studies have reported annual fluctuations in seizures, spiking in winter \citep{Unsal2020}. We also tested whether circadian properties differed across months of the year, finding substantial variation for both acrophase and amplitude, though period remained fixed year-round. Future work should explore this further and investigate the association between seasonal patterns of circadian variability and seizure frequency. 

%% SUMMARY: Other conditions
%Day-to-day variability in other conditions: depression, bi-polar

%% LIMITATIONS: Methodical
Our approach has some key limitations. Firstly, while heart rate follows a strong circadian cycle \citep{Vandewalle2007}, bouts of activity such as exercise or stress can cause `behavioural masking', interfering with measurement of the circadian state \citep{Cui2023}, particularly when derived from wearable PPG. Despite the issues of noise, resolution and accuracy associated with wearable devices, they provide key advantages in affordability, convenience and scalability over traditional methods for measuring the circadian rhythm such as melatonin sampling, enabling their wider use in clinical contexts and normative benchmarking. Future work should take advantage of multi-modal wearables, integrating heart rate with activity and temperature for circadian state estimation, for example. Secondly, our analysis does not include some key variables that may play a role in the intra-individual variability we observe. Anti-seizure medications (ASMs) have been reported to both stabilise and disrupt sleep \citep{Bazil2003, Derry2013}, and so could explain the increased intra-individual variability we see in PWE as well as the overlap with controls if this effect translates to the circadian rhythm more generally. Demographics such as age and sex are also known to associate with measures of the circadian rhythm \citep{Logan2019, Natarajan2020}. In a subset of the cohort (Supplementary~\ref{suppl:age_sex}), we did observe a correlation between age and intra-individual averages of acrophase and amplitude as expected based on this previous literature, but not with variability in any property. We did not find any sex differences with respect to circadian average or variability. Future work should further clarify whether associations with these potential co-variates can be ruled out.

% sleep: https://onlinelibrary.wiley.com/doi/full/10.1111/jsr.13967
%https://pmc.ncbi.nlm.nih.gov/articles/PMC8056704/
%https://link.springer.com/article/10.1007/s42761-021-00082-6

Finally, the role of sleep disruption in epilepsy is well noted \citep{Touchon1991, Hofstra2009, FoldvarySchaefer2006, Derry2013, Daley2018, Grayson2018, Bazhanova2022}, and while a distinct process physiologically \citep{Hofstra2009, Meyer2022, Vetter2018}, it is intrinsically tied to the circadian rhythm. As the circadian behaviour of heart rate is tied to sleep, its variability may primarily be driven by sleep disruptions. Future studies should incorporate sleep quality questionnaires and wearable sleep tracking to distinguish between sleep-related variability \textit{vs.} more broad intra-individual variability of the circadian rhythm.

%% OUTLOOK
%Future work could also develop a normative map of circadian variability across a diverse range of wearable modalities and demographic populations. Such a model would not only provide insight into the distribution of circadian variability in a wider healthy population, but would reduce the demands of similar clinical studies in regards to collection of control data. 

\section{Conclusion}

In conclusion, we found increased variability of the circadian rhythm of heart rate in epilepsy, which may be indicative of a pathological circadian rhythm disruption in epilepsy as detected using a commercial wearable device. The  effect driving this remains unclear; we were unable to detect any relationship with seizures, so we instead propose that co-morbidities, ASMs, sleep disruption or some other confounds may be involved, or that it is an additional effect - alongside seizures - of the cellular dysfunction underlying the aetiology found in some epilepsies. We hope our findings encourage further research of this phenomenon and whether it has application in seizure prediction and chronotherapy in epilepsy, and further encourage use of intra-individual variability of circadian properties as a wearable biomarker for disruption in other conditions.

\section{Data Availability}
% SEE LANCET DIGITAL HEALTH GUIDELINES FOR WAHT TO INCLUDE. Says should be right at end of paper. 
% thank participants
% melbourne data; on request to melbirne authors, but also share link from their paper on some public data

All wearable heart rate recordings and seizure metadata are available upon request to PK (\url{karoly.p@unimelb.edu.au}). A subset of this data is already publicly available on Figshare (\url{DOI 10.26188/15109896}) from a previous publication \citep{Karoly2021}. We would like to acknowledge and thank the participants of this study.  

\section{Code availability}

Analysis code is available on GitHub:
\\ \url{https://github.com/cnnp-lab/2024_Billy_Intra-individual_variability_circadian}.  

% \section{AI statement}
% During the preparation of this work the authors used ChatGPT in order to improve the flow of the text and revise for grammar and typos. After using this tool/service, the authors reviewed and edited the content again and take full responsibility for the content of the publication.

\section{Acknowledgements}
Data collection for the study was supported by the Australian Government National Health and Medical Research Council (Grant \#1178220). P.N.T. and Y.W. are both supported by UK Research \& Innovation (UKRI) Future Leaders Fellowships (MR/T04294X/1, MR/V026569/1).

\section{Author contributions}

 BCS contributed to conceptualization, methodology, software, formal analysis, writing and visualisation; CT contributed to methodology, software, writing, visualisation, and supervision; RES contributed to resources, data curation, validation, and writing; GMB and NE contributed to validation and supervision; SJG contributed to validation and methodology; PNT contributed to writing, visualisation, supervision; KL contributed to methodology, validation, writing and supervision;  PJK contributed to resources, data curation, validation, and writing; YW contributed to conceptualization, methodology, software, formal analysis, writing, visualisation, supervision, project administration, and funding acquisition. We also thank members of the Computational Neurology, Neuroscience \& Psychiatry Lab (\url{www.cnnp-lab.com}) for discussions on the methodology, analysis and manuscript.

\section{Conflict of interest}
None of the authors have any conflict of interest to disclose.
We confirm that we have read the Journal’s position on issues involved in ethical publication and affirm that this report is consistent with those guidelines.

\bibliography{circadian_variability}

%%%%%%%%%% %%%%%%%%%% SUPPLEMENTARY %%%%%%%%%% %%%%%%%%%% 

\section*{Figure legends}

\textbf{Figure 1:} An illustrative schematic of one `cycle' of the circadian rhythm of heart rate (CRHR) with the 3 descriptive properties used in this study labelled. Period reflects the duration of one cycle of the modelled CRHR (approximately 24 hours). Acrophase is defined as the time-of-day at peak cycle magnitude, reflecting rhythm ‘timing’.
Amplitude is a measure of circadian ‘strength’, defined as the difference in magnitude between cycle peak and previous trough. 

\textbf{Figure 2:} Participant heart rate data (A) was split into runs between $>$8h gaps of missing data (B). 
    SSA extracted the CRHR from each run (C to E(i)). 
    Troughs in the circadian phase (E(ii)) were used as reference points for splitting the extracted CRHR into individual cycles.
    Circadian properties were calculated for each cycle and grouped into segments of seven days (G).   
    The mean and standard deviation of each property was then calculated within each 7-day segment. 
    The mean across all segments in a subject produces the final intra-individual average and variability values. 
    For example, if this example participant had only these two segments, their intra-individual variability of acrophase  would be $(34 + 59)/2 = 46.5$ minutes. 

\textbf{Figure 3:} Comparison of the distribution of the intra-individual average and variability of circadian properties between PWE and controls. 
    Period, acrophase and amplitude are defined in Figure~\ref{fig:methods1}.
    Effect sizes are measured in Cohen's d and p-values are based on the two-sided Wilcoxon rank-sum test for comparison between PWE and Controls.
     Each subpanel has been zoomed-in to omit outliers; outliers were not excluded prior to statistical testing.

\textbf{Figure 4:} Scatter plots between intra-individual variability of circadian properties and seizure frequency.
        Seizure frequency is in units of seizures per week on log10 scale. 

\textbf{Figure 5:} Intra-individual average and variability of circadian properties calculated over segments with and without seizures.
     Controls are shown for reference and are not included in the statistics shown.
     The two-sided Wilcoxon signed-rank paired test was used for comparison of the intra-individual average and variability calculated using segments with and without seizures.
     Each subpanel has been zoomed-in to omit outliers; outliers were not excluded prior to statistical testing.

\renewcommand{\thefigure}{S\arabic{figure}}
\setcounter{figure}{0}
\counterwithin{figure}{section}
\counterwithin{table}{section}
\renewcommand\thesection{S\arabic{section}}
\setcounter{section}{0}

\newpage
\section*{Supplementary}

% \subsection*{Table of Contents}
% \begin{enumerate}
%     \item[\textbf{S1}] Methodological Details
%     \begin{enumerate}
%         \item[S1.1] Comparison of circadian rhythm extraction methods
%         \item[S1.2] Deriving circadian cycles for computing circadian properties
%         \item[S1.3] Accounting for gaps in recording and variability in recording du-
% ration between participants
%     \end{enumerate}
    
%     \item[\textbf{S2}] Recording duration does not affect intra-individual vari-
% ability estimates
%     \begin{enumerate}
%         \item[S2.1] The relationship between intra-individual variability of circadian properties and total
% duration.
%     \end{enumerate}

%     \item[\textbf{S3}] Results hold when we resample to account for unbal-
% anced samples
%     \begin{enumerate}
%         \item[S3.1] Overview of the random sub-sampling correction to Figure 3
%     \end{enumerate}

%     \item[\textbf{S4}] Increased intra-individual variability of circadian prop-
% erties is not age, sex or epilepsy type dependent
%     \begin{enumerate}
%         \item [S4.1] The relationship between intra-individual variability of circadian properties and averages
% and age.
%     \item[S4.2] Comparison of the distribution of the average and variability in each circadian
% property between PWE and controls, split by sex
%     \item[S4.3] Comparison of the distribution of the intra-individual average and variability
% of circadian properties between epilepsy types
    
%     \end{enumerate}
    
% \end{enumerate}

\section{Methodological details}\label{suppl:method_details}

\subsection{Comparison of circadian rhythm extraction methods}\label{suppl:modelling_methods_comparison}

Approaches for computational modelling of the circadian rhythm broadly fall into two categories: dynamic modelling, where the behaviour of the circadian system is represented within a set of equations, and statistical modelling, where a periodic function approximating an underlying core circadian fluctuation is fit to, or derived from, the data \citep{AsgariTarghi2019}. Here, we apply the latter statistical modelling methods. %Crucially, the descriptive parameters period, acrophase and amplitude can be derived (Figure \ref{fig:methods1}) from an approximated circadian rhythm containing at least one cycle.

The simplest statistical modelling approach involves the fitting of a sinusoid wave. The rhythm produced is fixed in period, acrophase and amplitude. However, it can often be empirically observed (especially in uncontrolled conditions) that these properties vary over time from one circadian cycle to the next \citep{AsgariTarghi2019}, potentially as the result of challenges to the circadian system (e.g inconsistent night shift working), variation in lifestyle and behaviour (e.g occasional late nights), or perhaps even health conditions (e.g the change between mood states in bipolar disorder, or seizure occurrence in epilepsy). 

Bandpass-filtering of the signal around 24 hours derives a circadian rhythm where amplitude is able to vary across cycles, but the other parameters remain somewhat fixed. More sophisticated methods exist, such as wavelet-based analysis (the continuous and discrete wavelet transform), signal decomposition techniques (such as empirical mode decomposition) and others such as singular spectrum analysis (SSA) (Figure \ref{fig:methods2}E(i)). An overview these methods can be found in \citep{Eriksen2023}. These methods are more flexible and produce a rhythm that fits the data better, capturing variation in all three properties across cycles.

\cite{GarciaIglesias2023} evaluated these as well as other methods for extraction of a circadian rhythm from a set of wearable measures (activity, heart rate, blood pressure, skin temperature, core temperature) recorded over one week from one individual. For each method and measure, they extracted a circadian rhythm and computed the goodness-of-fit ($R^2$) of the rhythm extracted compared to the original timeseries. While certain methods performed better on some measures, SSA was the best all-rounder across measures. We perform our own comparison using the same approach in Supplementary Figure~\ref{fig:goodness_of_fit}, though with a smaller subset of methods and only using heart rate, but crucially we test over a much larger cohort (31 controls and 143 PWE). SSA performs best across methods (average $R^2 \approx$ 0.4). Interestingly, there is a persistently lower $R^2$ across measures for PWE, potentially supportive of a more variable rhythm for PWE that is picked up poorly by less flexible methods (BANDPASS, COSINE), though could also be related to the increased sample size and recording duration of PWE - notably this difference is minimised for SSA. This finding further informs us that, on average, 40\% of the variability in heart rate timeseries was accounted for by the circadian rhythm in this study, the remaining 60\% associated with other factors, potentially exercise or acute stress.  

While \cite{GarciaIglesias2023} found that certain methods (EEMD, CEEMDAN) performed better in heart rate, the sample size of 1 individual should be noted, as our result shows that, between participants, there is considerable variation and overlap in the $R^2$ between methods. For this reason, as well as SSA's reliable performance in both \cite{GarciaIglesias2023} study and our own analysis, and our previous experience with these methods, we selected SSA for use in this study. Future work should compare a larger set of methods more thoroughly across a larger and more varied population.      

\begin{figure}
    \centering
    \includegraphics[width=1\linewidth]{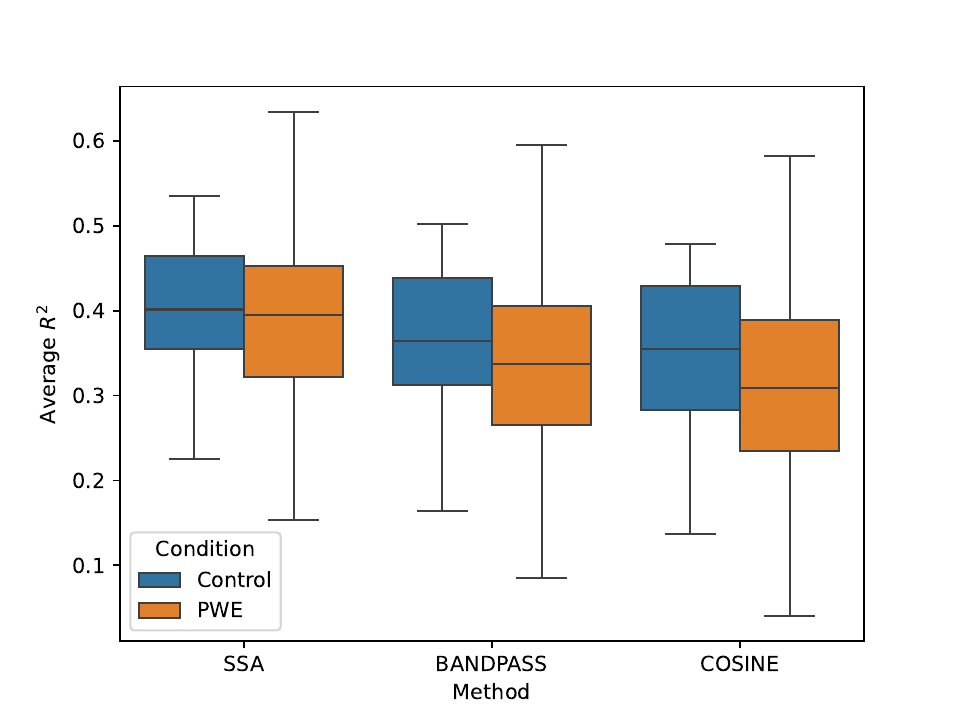}
    \caption{The distribution of goodness-of-fit ($R^2$ (derived from Pearson's \textit{r}) averaged across runs for each participant) of the circadian rhythm extracted by each method compared to the participant's original heart rate timeseries, split by PWE and controls. For methods requiring a minimum/maximum period range for the circadian rhythm the be detected within (CWT, BANDPASS), 20-30 hours were used respectively. This is the same threshold used to identify the circadian rhythm in the set of components extracted by SSA. }
    \label{fig:goodness_of_fit}
\end{figure}

\subsection{Deriving circadian cycles for computing circadian properties}\label{suppl:hilbert}
We split the rhythm into the individual circadian cycles (Figure \ref{fig:methods2}E(ii)), and compute the daily period, acrophase and amplitude  (Figure \ref{fig:methods2}G). To do so, the Hilbert transform is applied to the extracted circadian rhythm, producing a complex-valued analytic signal, from which the circadian ``phase series'' can be derived (Figure \ref{fig:methods2}E(ii)). The phase series is a periodic triangular waveform with bounds of $-\pi$ to $\pi$. It is aligned with the input circadian rhythm, and can be thought of as a measure of circadian progression; at $-\pi$ the rhythm is at trough, and reaches its peak as phase increases to 0. As phase increases further from 0 to $\pi$, the cycle falls again before it reaches the next trough at $\pi$, where it wraps back around to $-\pi$. Using this, we can robustly split the rhythm at each phase trough ($-\pi$) into the daily circadian cycles. Once we have collected individual cycles, the circadian properties (Figure \ref{fig:methods1}) are calculated for each cycle. \\

\subsection{Accounting for gaps in recording and variability in recording duration between participants}\label{supp:runs_and_segmenting_extra_detail}

As shown in Suppl. Table~\ref{tab:melbourne_data_after_processing}, the recording duration for both PWE and controls in this cohort is uniquely long, and there is substantial variation in recording duration between participants. These factors necessitated two additional steps in the calculation of intra-individual variability: the splitting of a participant's recording into ``runs'' between missing gaps prior to extraction of the circadian rhythm, and the grouping of extracted per-cycle circadian property values into ``segments'' prior to computation of mean and standard deviation. 

Missing data gaps in the recording were most likely caused by device charging or removal of the device by the participant for whatever other reason. As the heart rate recordings were computed by a proprietary Fitbit algorithm from raw PPG, it is possible that periods of noisy data (for example, due to poor device contact with skin) were excluded automatically also. SSA cannot handle missing data in the input recording, so interpolation (e.g linear) is required. However, there are occasionally very long (days-months) gaps of missing data for some participants, so we opted to avoid running the algorithm over very long interpolations of data, as this wasted computational resources and interfered with results. %Instead, we split the recording prior to extraction of the circadian rhythm into ``runs'' of data where any missing gaps are below 6 hours (Figure \ref{fig:methods2}B). Iterating over the recording, when a gap above 6 hours in duration is encountered, the run of valid data is ended and a new run begun when the valid data resumes. The 6 hour threshold was chosen as it corresponds to a quarter-turn of the circadian cycle, so linear interpolation should not drastically alter the extracted cycle. Additionally, the charging time for a Fitbit is well within this period (CITE? fitbit charging time). 
As such, the raw heart rate data was split into `runs' between missing gaps larger than 8 hours. Within each run, any remaining gaps (which as such must be below 8 hours) were linearly interpolated. \\

As stated previously, the intra-individual variability approach involves computation of a mean and standard deviation for each of the circadian properties (period, acrophase, amplitude) for each participant. However, given the variation in recording duration between participants (and especially between PWE and controls) in this dataset, and given that estimation of mean and standard deviation become more reliable as sample size increases, it would not be fair to compare the intra-individual variability between a participant with, for example, a recording duration of a month to a participant with multiple years worth of data. Therefore, once the circadian properties have been calculated for each cycle, rather than calculating the mean and standard deviation of each property across \textit{all} cycles, we group cycles into consecutive non-overlapping seven-day `segments'. \\

Grouping occurred within runs; for example, a run that contained 17 circadian cycles would produce 2 segments (14 cycles), with 3 cycles discarded. We opted to perform the segmenting within runs rather than across the cycles of all runs as, given the minimum 8 hour gap between runs, there may be a considerable gap of time between two consecutive runs.  \\

For each seven-day segment, the standard deviation of each property (Figure~\ref{fig:methods2}G) was calculated, reflecting the variability of that property during the corresponding week. Therefore, for each participant, a distribution of standard deviation values is produced for each property, reflecting differences in variability across weeks. To summarise these distributions for each participant, three summary values - intra-individual circadian period variability, acrophase variability and amplitude variability - were calculated by taking the mean of the standard deviation values of each property across the segments distribution. Three additional summary values: intra-individual period average, acrophase average and amplitude average, were calculated in a similar manner, using the mean of each property across all segments. \\

Calculating the mean and standard deviation separately for each segment ensures that they are always calculated on samples of data of the same length (7 cycles), reducing the problem of varying recording duration between participants. Taking the mean across all segments allows for comparison of circadian average and variability between participants. \\

%Admittedly, we would ideally use a minimum far greater than 7, however we found that the dropoff in participants retained was too great at thresholds greater than this (e.g 14 (2 weeks), 30 (approximately a month) - see Supplementary Figure ?). 

A visualisation of our implementation of the intra-individual variability method, with the addition of the segmenting step, can be found in Figure \ref{fig:methods2}.

\section{Demographic and clinical information table}

\begin{table}[!ht]
    \centering
    \begin{tabular}{|l|l|l|l|l|}
    \hline
        \textbf{} & \textbf{Controls} & \textbf{Epilepsy} & \textbf{Statistic} & \textbf{p-value} \\ \hline
        \textbf{N} & 31 & 143 & ~ & ~ \\ \hline
        \textbf{Age} mean (years) & 37.0 & 38.6 & \multirow{3}{*}{-0.215} & \multirow{3}{*}{0.84} \\  \cline{1-3}
        \textbf{Age} sd (years) & 16.9 & 13.3 & ~ & ~ \\ \cline{1-3}
        \textbf{Age} unavailable & 26 & 39 & ~ & ~ \\ \hline  
        \textbf{Sex} female & 15 & 87 & \multirow{4}{*}{2.777} & \multirow{4}{*}{0.427} \\  \cline{1-3}
        \textbf{Sex} male & 13 & 39 & ~ & ~ \\  \cline{1-3}
        \textbf{Sex} other & 0 & 1 & ~ & ~ \\  \cline{1-3}
        \textbf{Sex} unavailable & 3 & 16 & ~ & ~ \\ \hline
        \textbf{Recording Duration} mean (days) & 221.5 & 443.9 & \multirow{5}{*}{1686} & \multirow{5}{*}{\textbf{0.037}} \\  \cline{1-3}
        \textbf{Recording Duration} sd (days) & 256.4 & 531.1 & ~ & ~ \\ \cline{1-3}
        \textbf{Recording Duration} median (days) & 125.1 & 215.8  & ~ & ~ \\ \cline{1-3}
        \textbf{Recording Duration} IQR (days) & 260.5  & 580.9 & ~ & ~ \\ \cline{1-3}
        \textbf{Recording Duration} (min, max) (days) & (7.0, 946.0) & (6.7, 2715.6)  & ~ & ~ \\ \hline
        \textbf{Focal} & ~ & 41 & ~ & ~ \\  \cline{1-3}
        \textbf{Generalised} & ~ & 13 & ~ & ~ \\  \cline{1-3}
        \textbf{Mixed} & ~ & 3 & ~ & ~ \\  \cline{1-3}
        \textbf{Total Number of Seizures} & ~ & 6110 & ~ & ~ \\ \cline{1-3}
        \textbf{Average Number of Seizures} & ~ & 43.4 & ~ & ~ \\ \hline
     \end{tabular}
     \caption{Demographic and clinical characteristics of the cohort. The distribution of age was compared between PWE and controls using the independent two-sided Welch's t-Test. Sex comparison was performed using the Chi-square test of independence. Duration comparison was performed using the two-sided Mann-Whitney U test.}
    \label{tab:melbourne_data_after_processing}
\end{table}

\section{Recording duration does not affect intra-individual variability estimates }\label{suppl:rec_duration}

\begin{figure}
    \centering
    \includegraphics[width=1\linewidth]{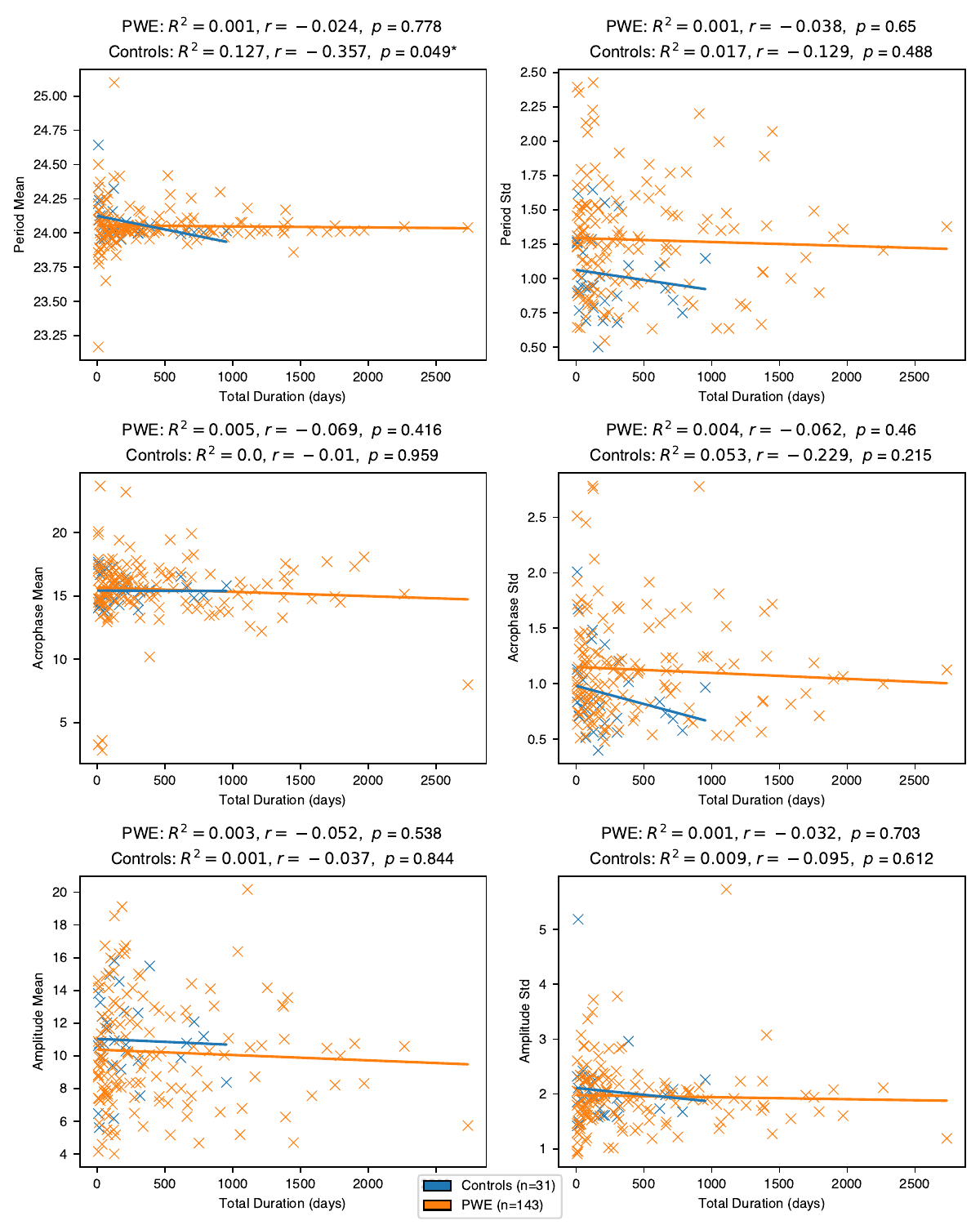}
    \caption{The relationship between intra-individual variability of circadian properties and total duration. }
    \label{fig:rec_duration}
\end{figure}

\section{Results hold when we resample to account for unbalanced samples }\label{suppl:subsampling}
%https://tex.stackexchange.com/questions/271518/multiple-panel-figure-with-figures-side-by-side
%https://tex.stackexchange.com/questions/302464/combining-tables-and-figures-with-common-subtitle
\begin{figure}[h]
    
    \centering
    \begin{subfigure}[t]{0.49\textwidth}
        \centering
        \includegraphics[width=\linewidth]{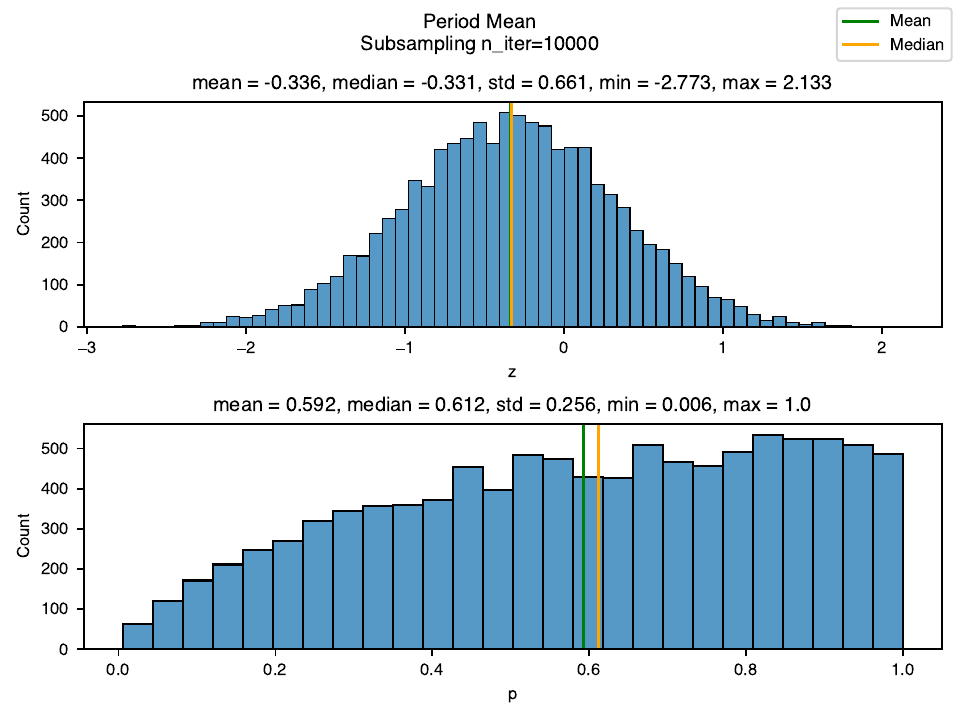} 
        \caption{Period Mean}
        \label{fig:timing1}
    \end{subfigure}
    \hfill
    \begin{subfigure}[t]{0.49\textwidth}
        \centering
        \includegraphics[width=\linewidth]{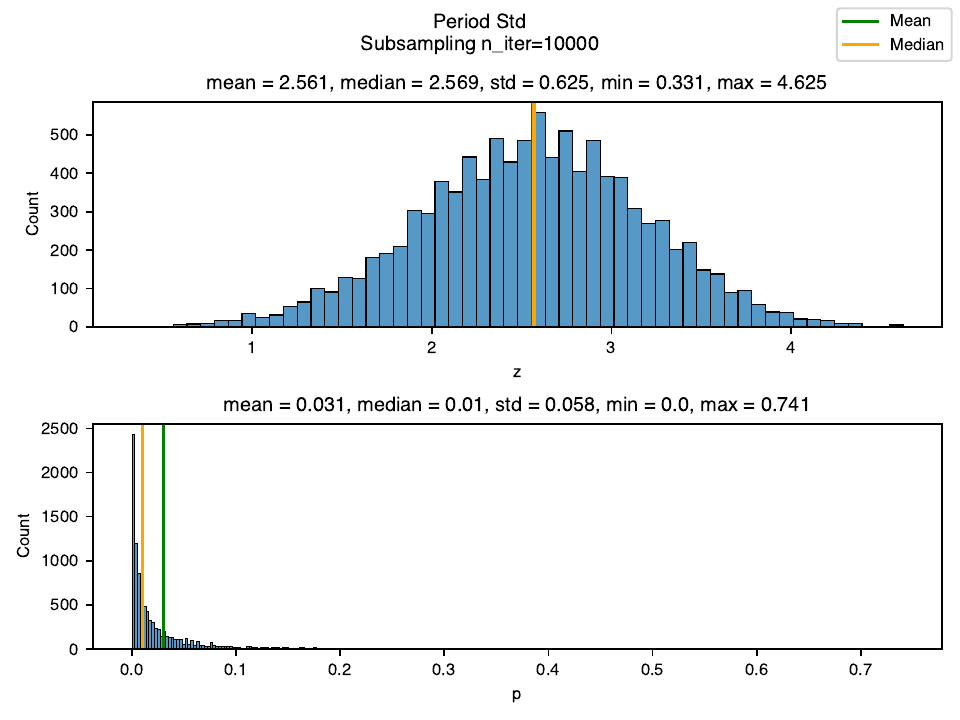} 
        \caption{Period Std}
        \label{fig:timing2}
    \end{subfigure}

    % \vspace{1cm}
    % \begin{subfigure}[t]{\textwidth}
    % \centering
    %     \includegraphics[width=\linewidth]{Period Std_subsampling.pdf} 
    %     \caption{Price regulation} \label{fig:timing3}
    % \end{subfigure}
    % \caption{Some general caption of all the figures. In (\subref{fig:timing1}) you can see a  green square....}

    \vspace{1cm}
    \begin{subfigure}[t]{\textwidth}
    \centering
        \begin{tabular}{|l|l|l|}
        \hline
            ~ & median z & median p \\ \hline
            Period Mean & -0.331 & 0.612 \\ \hline
            Period Std & 2.569 & \textbf{0.01} \\ \hline
            Acrophase Mean & 0.739 & 0.435 \\ \hline
            Acrophase Std & 2.231 & \textbf{0.026} \\ \hline
            Amplitude Mean & -1.19 & 0.234 \\ \hline
            Amplitude Std & -0.528 & 0.531 \\ \hline
        \end{tabular}
        \caption{Table of random sub-sampling median p-values and z-statistics} \label{fig:timing3}
    \end{subfigure}
    
    \caption{\textbf{Overview of the random sub-sampling correction to Figure \ref{fig:results1}.} (a) and (b) show the distribution of the z-statistic and p-value over 10,000 iterations of the Wilcoxon rank-sum test using the entire Control cohort and a randomly selected sub-sample of 31 PWE. (a) shows this distribution for Period Mean (intra-individual average of period), which was not originally different between PWE and controls. This result holds after applying random sub-sampling as the z-statistic is normally distributed around $\sim$0 (no effect) and the p-value distribution is uniform. (b) shows this distribution for Period Std (intra-individual variability of period), which was originally different between PWE and controls. This result also holds after applying random sub-sampling, as the z-statistic is normally distributed around 2.5 (moderate effect) and the p-value distribution is extremely skewed with a median $<$ 0.05. (c) shows the median z-statistic and p-values across each property average and variability. }
    
\end{figure}

\section{Increased intra-individual variability of circadian properties is not age, sex or epilepsy type dependent}\label{suppl:age_sex}

\begin{figure}
    \centering
    \includegraphics[width=1\linewidth]{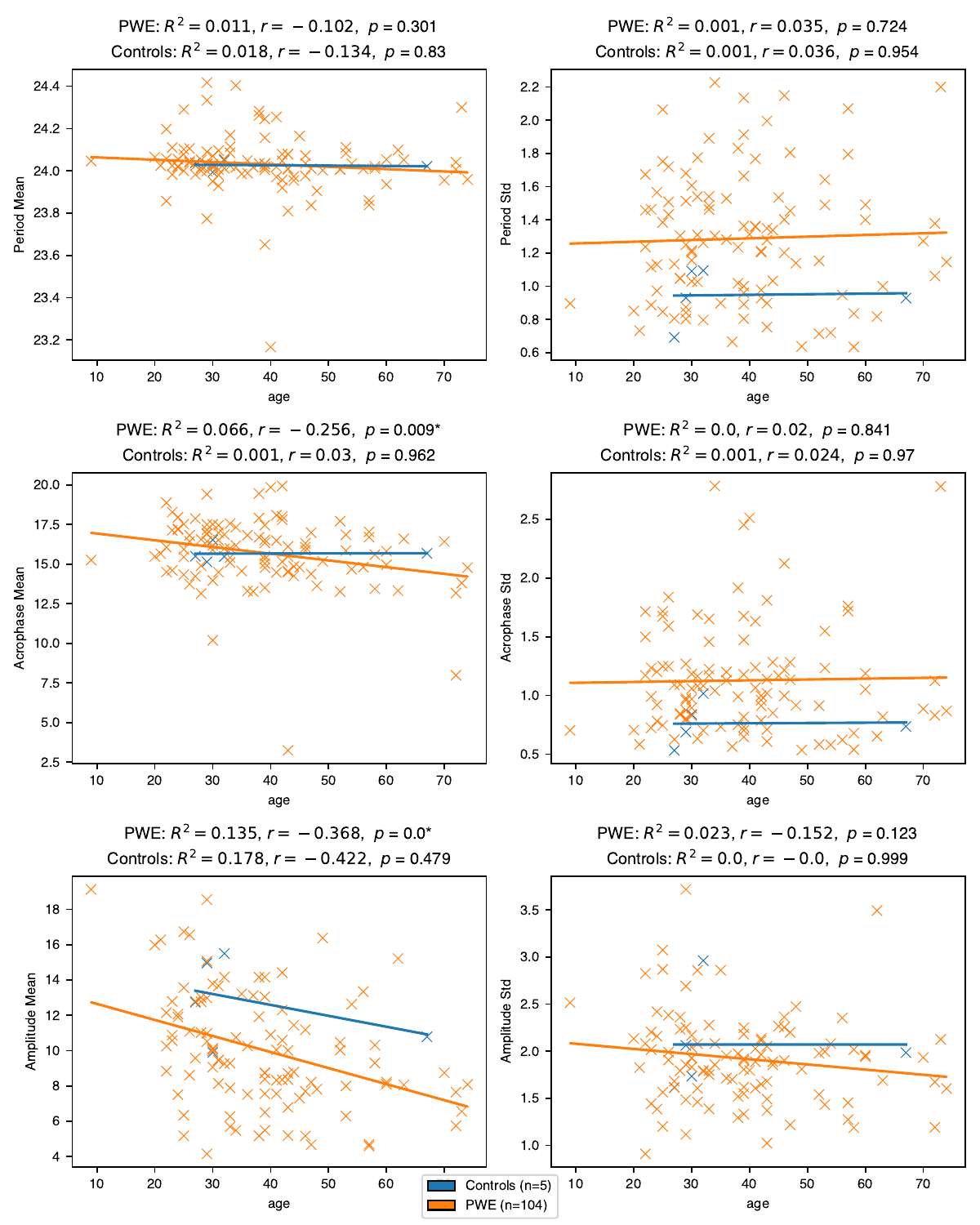}
    \caption{The relationship between intra-individual variability of circadian properties and averages and age.}
    \label{fig:age}
\end{figure}

\begin{figure}
    \centering
    \includegraphics[width=1\linewidth]{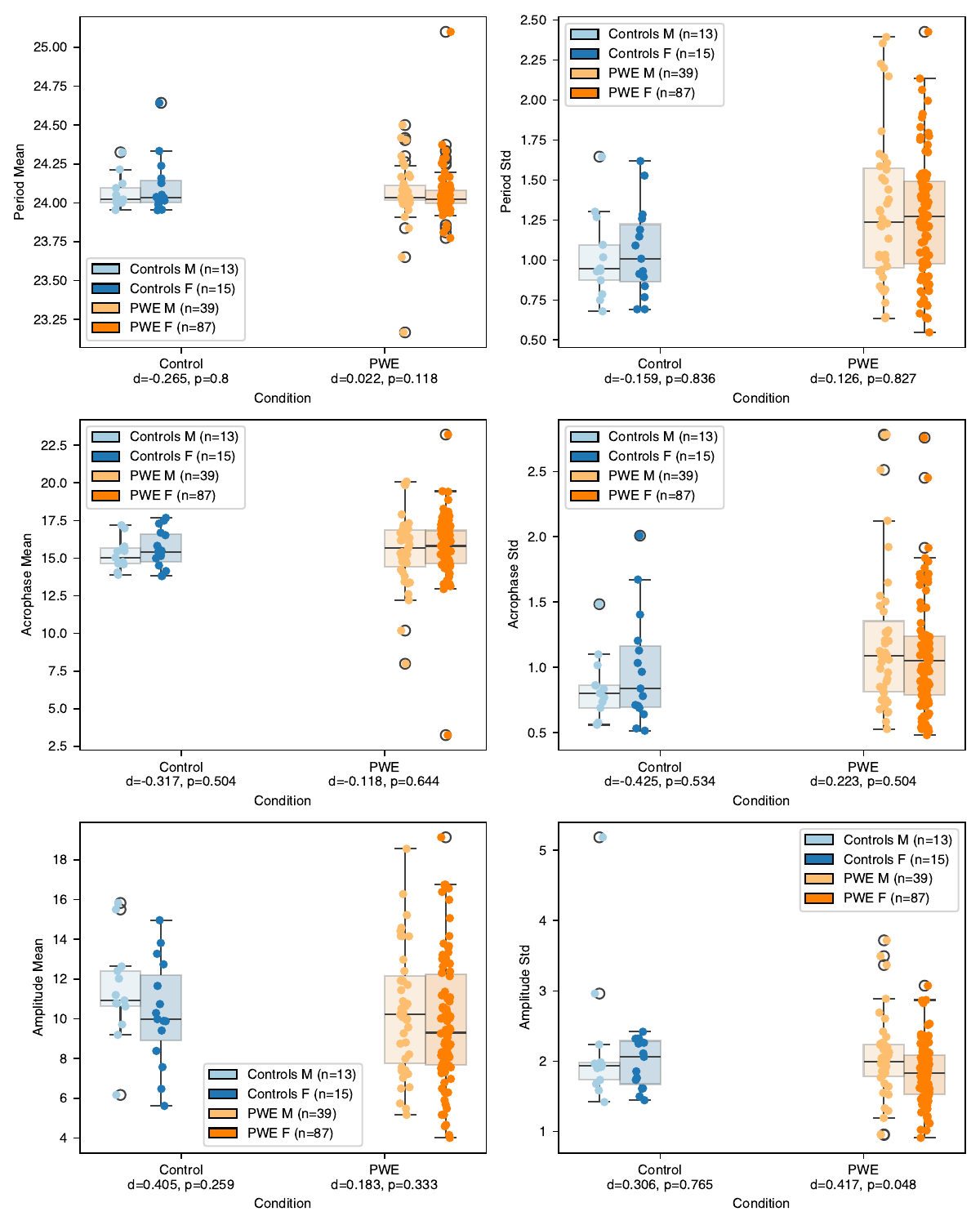}
    \caption{\textbf{Comparison of the distribution of the average and variability in each circadian
property between PWE and controls, split by sex.} Two-sided Wilcoxon rank-sum test used to compare circadian average and variability between males and females within either PWE or controls. }
    \label{fig:sex}
\end{figure}

\begin{figure}
    \centering
    \includegraphics[width=1\linewidth]{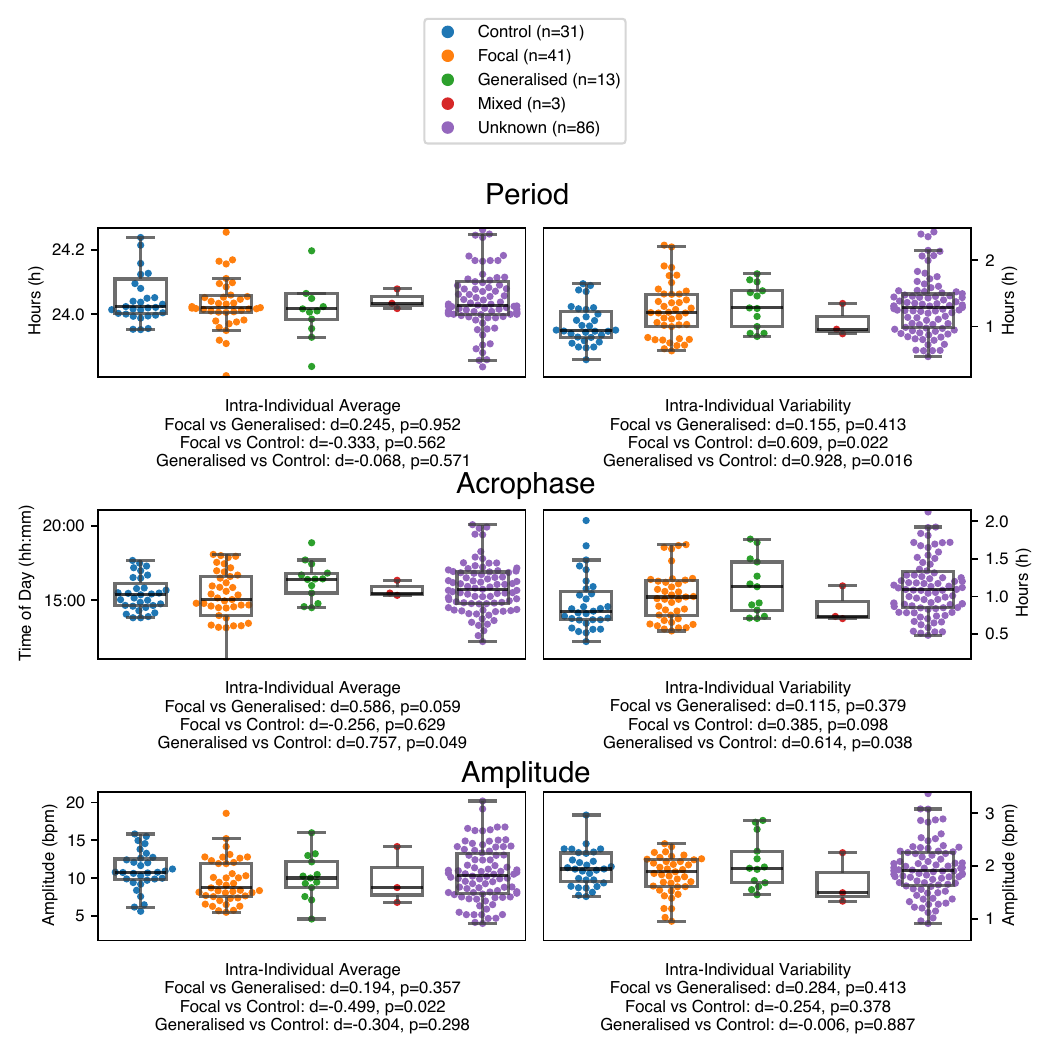}
    \caption{\textbf{Comparison of the distribution of the intra-individual average and variability of circadian properties between epilepsy types.}
     Mixed type and unknown (not provided) type PWE are shown for reference only and are not included in the statistics shown.
     The two-sided Wilcoxon rank-sum test was used for comparison of the intra-individual average and variability between generalised and focal epilepsy, and each sub-type against controls.
     Each subpanel has been zoomed-in to omit outliers; outliers were not excluded prior to statistical testing.
     }
    \label{fig:epilepsy_type}
\end{figure}

%\section{Increased intra-individual variability of circadian properties is consistent across epilepsy type}\label{suppl:epilepsy_type}

%\section{Circadian property mean values do not themselves relate to seizure occurrence \label{suppl:mean_sz}}
% this was the results figure 3 presented at paper pitch
% I think its worth checking as data is there, and there is intuition that on subject specific level, maybe certain abnormal circadian values drive seizure occurrence. But finding is not that interesting compared to others, and is confusing story-wise as we're going back a step to before we calculate mean and standard deviation
% i think this would be correlatng mean (not sd) properties of a cycle themeselves with either whether seizures occurred on that day or not, or number of seizures

\section{Impact of weekends and time of year on circadian properties and their intra-individual average and variability}\label{suppl:weekly_annual}

To investigate whether variability in the circadian rhythm is associated with weekday \textit{vs.} weekend effects or seasonal variations, circadian cycles (Figure \ref{fig:methods2}G) were associated with a date (at acrophase), and 7-day segments were associated with their `time of year' (see below). 

\subsection{Weekday vs weekend effects}\label{suppl:weekly}

% INTRO
`Social jetlag' refers to the phenomenon where sleep patterns differ on weekdays compared to weekends due to the constraints of work, education or other commitments \citep{Vetter2018}. More pronounced social jetlag has been reported for PWE compared to controls \cite{Choi2016}. Sleep patterns, while distinct from the circadian rhythm, are linked. As such, we test here whether circadian variability is similarly associated with weekday/weekend differences, and whether this varies between PWE and controls. 

\begin{figure}
    \centering
    \includegraphics[width=0.8\linewidth]{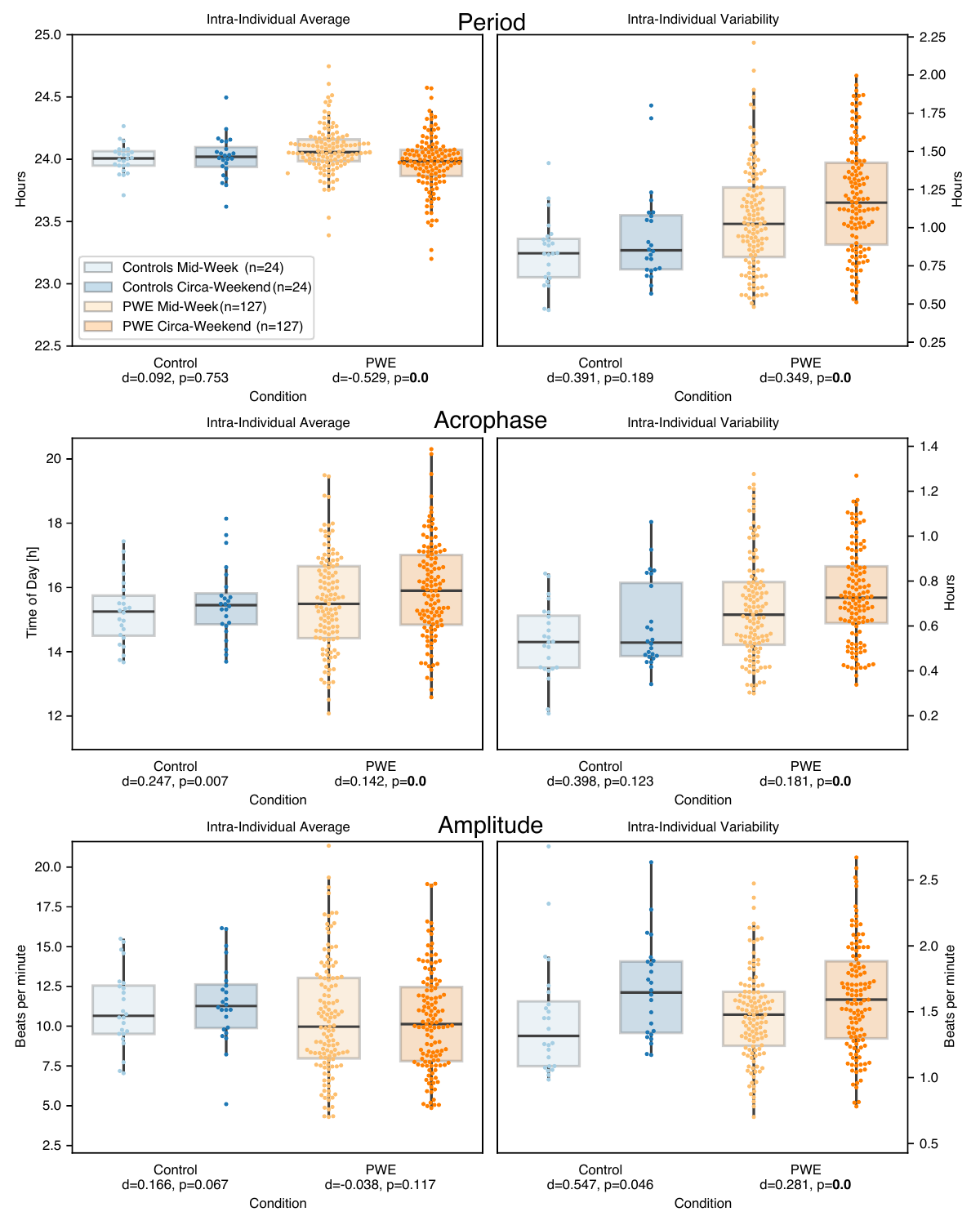}
    \caption{Intra-individual average and variability of circadian properties calculated over mid-week (Tue, Wed, Thu, lighter colours) and circa-weekend (Fri, Sat, Sun, darker colours) segments separately for both PWE and controls.
     The two-sided Wilcoxon signed-rank paired test was performed for comparison of the intra-individual average and variability calculated using mid-week vs circa-weekend segments for PWE and controls independently (i.e, \textbf{is there a weekend effect within each group?}). Only participants with at least 3 mid-week and at least 3 circa-weekend segments were kept. Highlighted p-values fall below the Bonferroni-corrected alpha threshold of 0.05/12 = 0.0042.}
     \label{fig:supp_weekend_within_group}
\end{figure}

\begin{figure}
    \centering
    \includegraphics[width=0.8\linewidth]{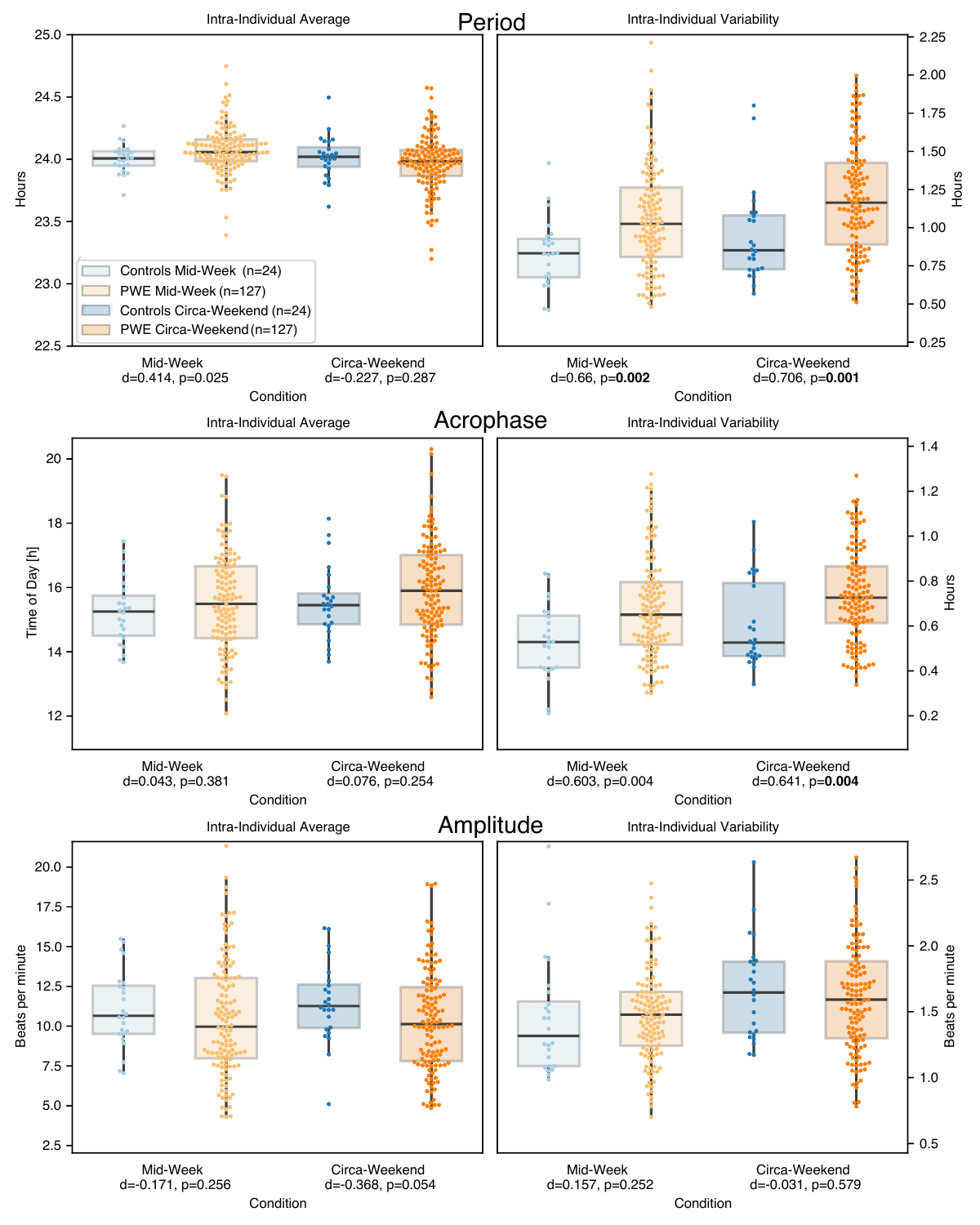}
    \caption{Intra-individual average and variability of circadian properties calculated over mid-week (Tue, Wed, Thu, lighter colours) and circa-weekend (Fri, Sat, Sun, darker colours) segments separately for both PWE and controls.
     The two-sided Wilcoxon rank-sum test was performed for comparison of the intra-individual average and variability calculated using mid-week vs circa-weekend segments separately, between controls and PWE (i.e, \textbf{is there a weekend effect between groups?}). Only participants with at least 3 mid-week and at least 3 circa-weekend segments were kept. Highlighted p-values fall below the Bonferroni-corrected alpha threshold of 0.05/12 = 0.0042. }
     \label{fig:supp_weekend_between_groups}
\end{figure}

% METHODS
To compare intra-individual circadian average and variability between weekends and weekdays, each segment was split into `mid-week' (Tue, Wed, Thu) and `circa-weekend' (Fri, Sat, Sun) sub-segments, and intra-individual circadian average and variability values were calculated over mid-week and circa-weekend segments separately. The equal 3-day segment size was selected as a fully weekday/weekend (Mon-Fri vs Sat-Sun) comparison introduces a sample size bias (5 vs 2) that interferes with mean and standard deviation interpretation. Furthermore, we hypothesize this may account for behavioural shifts - individuals who experience social jetlag are most likely adjusted to their weekday schedule by Tuesday (following a disruptive Monday), and there is maybe a shift in behaviour on Friday evening compared to the evenings on days prior. Firstly, intra-individual circadian average and variability is compared between weekends and weekdays within PWE and control groups separately in Figure~\ref{fig:supp_weekend_within_group} to determine if a `weekend effect' exists for both groups or is unique to either. Following this, Figure~\ref{fig:supp_weekend_between_groups} compares intra-individual circadian average and variability between PWE and controls on weekdays and weekends separately, to determine if previously observed differences in these properties between PWE controls are consistent across the week, or are being driven by a weekend effect. 

% IIV weekend figure

\newpage

To further explore the association between intra-individual average and variability of circadian properties in epilepsy and the weekend effect and determine the relative importance of these factors, a mixed-effects linear regression (MELR) model was used:
\begin{equation}
    \begin{split}
    property\_measure \sim & total\_duration\_d \\
    & + condition \\
    & + weekend \\
    & + (condition:weekend) \\
    & + (1|subject)  
    \end{split}
    \label{eqn:weekend_MELR_full}
\end{equation}

Where $property\_measure$ refers to Period Mean, Period Std, Acrophase Mean, etc, $total\_duration\_d$ refers to the total duration in days of all mid-week or circa-weekend segments for an individual, $condition$ refers to whether the individual is a control or PWE, $weekend$ refers to whether the $property\_measure$ is calculated over the mid-week (False) or circa-weekend (True) segments, and $(condition:weekend)$ refers to the interaction between the $condition$ and $weekend$ terms. A random intercept was applied for each participant to account for systematic differences between participants that may arise due to chronotype or routines (work, exercise, etc). The outputs of this model are shown in Table~\ref{tbl:weekend_MELR} with p-values $<0.05$ highlighted. 

Additionally, to determine the relative importance of modelling the weekend effect, we performed likelihood ratio tests between the `full' model (shown in Equation~\ref{eqn:weekend_MELR_full}) and a `reduced' model, that incorporates the weekend effect but not its interaction with condition (Equation~\ref{eqn:weekend_MELR_reduced}) and a `null' model that does not consider the weekend effect at all (Equation~\ref{eqn:weekend_MELR_null}). p-values from these tests are reported in Table~\ref{tbl:weekend_LR}, with p-values $<$ 0.05 highlighted.

\begin{equation}
    \begin{split}
    property\_measure \sim & total\_duration\_d \\
    & + condition \\
    & + weekend \\
    & + (1|subject)  
    \end{split}
    \label{eqn:weekend_MELR_reduced}
\end{equation}

\begin{equation}
    \begin{split}
    property\_measure \sim & total\_duration\_d \\
    & + condition \\
    & + (1|subject)  
    \end{split}
    \label{eqn:weekend_MELR_null}
\end{equation}

% RESULTS

In Figure \ref{fig:supp_weekend_within_group}, most intra-individual average and variability measures differ between weekdays and weekends in PWE and controls. We avoid direct comparison of effect sizes or p-values here, as there are substantially different sample sizes underlying the PWE \textit{vs.} controls cohorts. Instead we report them in Figure~\ref{fig:supp_weekend_within_group} ``as is''.

In Figure \ref{fig:supp_weekend_between_groups}, for each variability measure, there is a substantial difference in intra-individual variability of period and acrophase (Cohen's d $\geq0.6$), regardless of weekend or weekday. This confirms that the weekend effects observed in Fig~\ref{fig:supp_weekend_within_group} are not driving the results observed in Fig~\ref{fig:results1}.

The MELR models outputs are reported in Table~\ref{tbl:weekend_MELR}. Using the models to evaluate the interaction between the weekend and condition effects, there is weak evidence that intra-individual averages in Period and Amplitude are reduced on weekends only for PWE, although the associated p-values would not survive a correction for multiple comparison. No evidence of an interaction effect was seen in any of the intra-individual variability measures.

In Table~\ref{tbl:weekend_LR}, only Period and Amplitude Means have p-values below 0.05 in the reduced column, suggesting modelling the interaction between weekend and condition only can improve description of the data for these variables, which aligns with their entries in Table \ref{tbl:weekend_MELR}. However incorporation of at least the weekend effect alone is clearly necessary, as all p-values are $<<$ 0.05 when comparing the null model (no incorporation of weekend effects) to the full.

% DISCUSSION
Intra-individual circadian average and variability values appear to differ between weekdays and weekends for PWE and controls; but importantly, these differences do not drive the increased circadian variability we report in the main text. The weekday vs. weekend effect appear independently for both controls and PWE.

\begin{table}

    \tiny
    
    \centering
   \begin{tabularx}{\textwidth}{|X|l|l|l|l|l|l|}
    \hline
        \textbf{Period Mean} & Coef. & Std.Err. & z & $P>|z|$ & [0.025 & 0.975] \\ \hline
        Intercept & 24.005 & 0.042 & 570.531 & \textbf{0.000} & 23.922 & 24.087 \\ \hline
        C(Condition)[T.PWE] & 0.073 & 0.045 & 1.607 & 0.108 & -0.016 & 0.162 \\ \hline
        C(Weekend)[T.True] & 0.013 & 0.055 & 0.243 & 0.808 & -0.095 & 0.122 \\ \hline
        C(Condition)[T.PWE]:C(Weekend)[T.True] & -0.122 & 0.060 & -2.027 & \textbf{0.043} & -0.240 & -0.004 \\ \hline
        total\_duration\_d & -0.000 & 0.000 & -0.019 & 0.985 & -0.000 & 0.000 \\ \hline
        Subject Var & 0.004 & ~ & ~ & ~ & ~ & ~ \\ \hline
    \end{tabularx}
    \vspace{5mm}

    \centering
    \begin{tabularx}{\textwidth}{|X|l|l|l|l|l|l|}
    \hline
        \textbf{Period Std} & Coef. & Std.Err. & z & $P>|z|$ & [0.025 & 0.975] \\ \hline
        Intercept & 0.860 & 0.072 & 11.994 & \textbf{0.000} & 0.719 & 1.000 \\ \hline
        C(Condition)[T.PWE] & 0.240 & 0.077 & 3.119 & \textbf{0.002} & 0.089 & 0.391 \\ \hline
        C(Weekend)[T.True] & 0.105 & 0.062 & 1.712 & 0.087 & -0.015 & 0.226 \\ \hline
        C(Condition)[T.PWE]:C(Weekend)[T.True] & 0.019 & 0.067 & 0.290 & 0.772 & -0.112 & 0.151 \\ \hline
        total\_duration\_d & -0.000 & 0.000 & -1.364 & 0.173 & -0.000 & 0.000 \\ \hline
        Subject Var & 0.072 & 0.073 & ~ & ~ & ~ & ~ \\ \hline
    \end{tabularx}
    \vspace{5mm}

    \begin{tabularx}{\textwidth}{|X|l|l|l|l|l|l|}
    \hline
        \textbf{Acrophase Mean} & Coef. & Std.Err. & z & $P>|z|$ & [0.025 & 0.975] \\ \hline
        Intercept & 15.312 & 0.470 & 32.555 & \textbf{0.000} & 14.390 & 16.233 \\ \hline
        C(Condition)[T.PWE] & 0.137 & 0.504 & 0.273 & 0.785 & -0.850 & 1.125 \\ \hline
        C(Weekend)[T.True] & 0.265 & 0.113 & 2.349 & \textbf{0.019} & 0.044 & 0.487 \\ \hline
        C(Condition)[T.PWE]:C(Weekend)[T.True] & 0.076 & 0.123 & 0.621 & 0.535 & -0.165 & 0.318 \\ \hline
        total\_duration\_d & -0.001 & 0.001 & -0.635 & 0.525 & -0.002 & 0.001 \\ \hline
        Subject Var & 4.884 & 2.063 & ~ & ~ & ~ & ~ \\ \hline
    \end{tabularx}
    \vspace{5mm}

    \begin{tabularx}{\textwidth}{|X|l|l|l|l|l|l|}
    \hline
        \textbf{Acrophase Std} & Coef. & Std.Err. & z & $P>|z|$ & [0.025 & 0.975] \\ \hline
        Intercept & 0.550 & 0.052 & 10.651 & \textbf{0.000} & 0.449 & 0.651 \\ \hline
        C(Condition)[T.PWE] & 0.178 & 0.055 & 3.204 & \textbf{0.001} & 0.069 & 0.286 \\ \hline
        C(Weekend)[T.True] & 0.072 & 0.046 & 1.555 & 0.120 & -0.019 & 0.162 \\ \hline
        C(Condition)[T.PWE]:C(Weekend)[T.True] & -0.025 & 0.050 & -0.494 & 0.621 & -0.123 & 0.074 \\ \hline
        total\_duration\_d & -0.000 & 0.000 & -1.645 & 0.100 & -0.000 & 0.000 \\ \hline
        Subject Var & 0.036 & 0.049 & ~ & ~ & ~ & ~ \\ \hline
    \end{tabularx}
    \vspace{5mm}

     \begin{tabularx}{\textwidth}{|X|l|l|l|l|l|l|}
    \hline
        \textbf{Amplitude Mean} & Coef. & Std.Err. & z & $P>|z|$ & [0.025 & 0.975] \\ \hline
        Intercept & 11.153 & 0.683 & 16.342 & \textbf{0.000} & 9.816 & 12.491 \\ \hline
        C(Condition)[T.PWE] & -0.549 & 0.731 & -0.751 & 0.453 & -1.981 & 0.884 \\ \hline
        C(Weekend)[T.True] & 0.411 & 0.236 & 1.742 & 0.082 & -0.051 & 0.873 \\ \hline
        C(Condition)[T.PWE]:C(Weekend)[T.True] & -0.543 & 0.257 & -2.114 & \textbf{0.035} & -1.047 & -0.040 \\ \hline
        total\_duration\_d & -0.001 & 0.001 & -0.460 & 0.645 & -0.003 & 0.002 \\ \hline
        Subject Var & 9.928 & 2.045 & ~ & ~ & ~ & ~ \\ \hline
    \end{tabularx}
    \vspace{5mm}

    \begin{tabularx}{\textwidth}{|X|l|l|l|l|l|l|}
    \hline
        \textbf{Amplitude Std} & Coef. & Std.Err. & z & $P>|z|$ & [0.025 & 0.975] \\ \hline
        Intercept & 1.459 & 0.099 & 14.730 & \textbf{0.000} & 1.265 & 1.653 \\ \hline
        C(Condition)[T.PWE] & 0.079 & 0.106 & 0.748 & 0.455 & -0.129 & 0.288 \\ \hline
        C(Weekend)[T.True] & 0.222 & 0.080 & 2.790 & \textbf{0.005} & 0.066 & 0.379 \\ \hline
        C(Condition)[T.PWE]:C(Weekend)[T.True] & -0.085 & 0.087 & -0.982 & 0.326 & -0.256 & 0.085 \\ \hline
        total\_duration\_d & -0.000 & 0.000 & -0.814 & 0.415 & -0.001 & 0.000 \\ \hline
        Subject Var & 0.148 & 0.110 & ~ & ~ & ~ & ~ \\ \hline
    \end{tabularx}

    \caption{Summary of the mixed effects models produced by Equation~\ref{eqn:weekend_MELR_full}.} 
    \label{tbl:weekend_MELR}
\end{table}

\begin{table}[!ht]
    \centering
    \begin{tabular}{|l|l|l|}
    \hline
        ~ & reduced & null \\ \hline
        Period Mean & \textbf{0.032} & \textbf{0.0} \\ \hline
        Period Std & 0.772 & \textbf{0.0} \\ \hline
        Acrophase Mean & 0.535 & \textbf{0.0} \\ \hline
        Acrophase Std & 0.621 & \textbf{0.021} \\ \hline
        Amplitude Mean & \textbf{0.036} & 0.098 \\ \hline
        Amplitude Std & 0.327 & \textbf{0.0} \\ \hline
    \end{tabular}
    \caption{p-values for likelihood ratio tests comparing reduced (Equation~\ref{eqn:weekend_MELR_reduced}) and null (Equation~\ref{eqn:weekend_MELR_null}) models to the full (Equation~\ref{eqn:weekend_MELR_full}) model for each property measure. p $<$ 0.05 reject the null hypothesis and suggests the full model fits the data significantly better and should be used. }
    \label{tbl:weekend_LR}
\end{table}

\newpage
\subsection{Time of year effects}\label{suppl:annual}

% INTRO

Seizures have been shown to exhibit seasonal variation, possibly clustering in winter \citep{Unsal2020}, and seizure occurrences have not only been found to be phase-locked to circadian rhythms of heart rate for some PWE, but longer-term (weekly and monthly) rhythms of heart rate also \citep{Karoly2021}. Here, we test whether intra-individual variability is associated with long-term seasonal trends, and whether this varies between PWE and controls.

\begin{figure}
    \centering
    \includegraphics[width=0.5\linewidth]{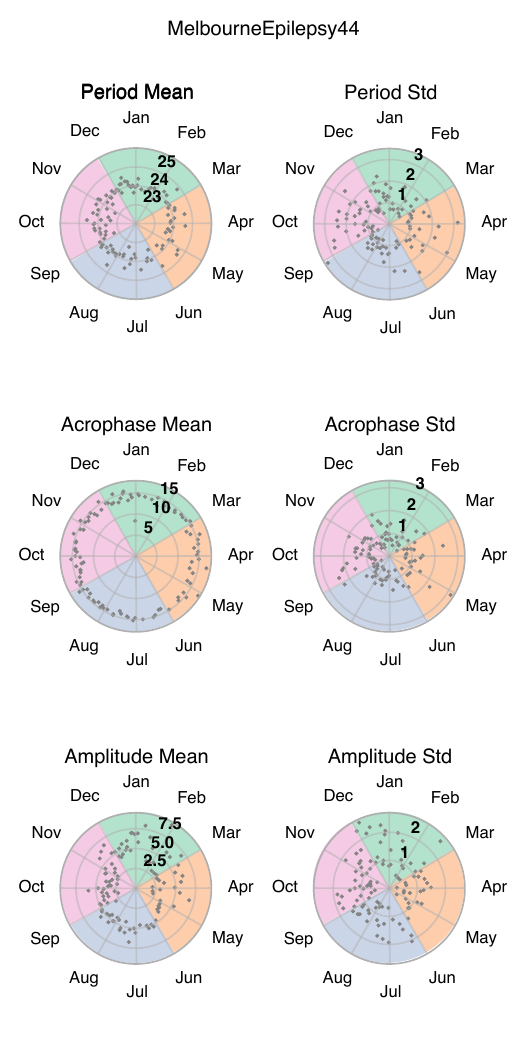}
    \caption{A visualisation of the data used in the MELR for one example participant. Each point corresponds to the circadian average or variability value for one 7-day segment, placed corresponding to its position within the year (data may span multiple years). Acrophase means have been corrected for local timezone and daylight saving time shifts. Coloured regions correspond to the seasons (Australian): Spring (Sept-Nov, pink), Summer (Dec-Feb, green), Autumn (Mar-May, orange), and Winter (Jun-Aug, blue). }
    \label{fig:supp_annual_cycles_polar}
\end{figure}

%METHODS
To explore changes in intra-individual circadian average and variability over the year, 7-day segments (Figure \ref{fig:methods2}G) were associated with the `time of the year' ($toy$) at which they occurred. This is defined as $toy = \frac{day\_of\_year}{days\_in\_year}$ where $day\_of\_year$ is an integer 1-365 (e.g. 1=1st January, 365=31st December) corresponding to the 4th (middle) circadian cycle of each 7-day segment, and days\_in\_year is 366 on leap years or 365 on regular years. Rather than producing a summary statistic for each participant by calculating the mean of circadian average and variability values across all segments as before, we can investigate whether these individual segment values vary depending on the time of year of the segment. This allows for a higher resolution in time along the year, and this is visualised for an example participant in Figure \ref{fig:supp_annual_cycles_polar}.

\newpage

To model the association between intra-individual average and variability of circadian properties in epilepsy and annual variations and determine the relative importance of these factors, a mixed-effects linear regression model (MELR) was used:

\begin{equation}
    \begin{split}
    property\_measure \sim & condition \\
    & + sin(toy\_radians) \\
    & + cos(toy\_radians) \\
    & + (condition:(sin(toy\_radians) + cos(toy\_radians))) \\
    & + (1|subject)  
    \end{split}
    \label{eqn:monthly_MELR_full}
\end{equation}

Where $property\_measure$ refers to Period Mean, Period Std, Acrophase Mean, etc, of a given segment, and $condition$ refers to whether the individual is a control or PWE. As the time of year is a circular variable, it cannot be used directly in a linear regression model. To account for this, \textit{cos} and \textit{sin} of the month variable (after conversion to radians ($toy\_radians = toy \cdot 2\pi$)) were included. A significant association with either one or both of these terms can be interpreted as a seasonal effect being present. The interaction between these terms and condition was added via $condition:(sin(toy\_radians) + cos(toy\_radians))$. A random intercept was applied for each participant to account for systematic differences between participants that may arise due to chronotype or routines (work, exercise, etc); but here, also to account for different number of 7-day segments between subjects. The outputs of this model are shown in Table~\ref{tbl:monthly_MELR} with p-values $<$ 0.05 highlighted. 

Additionally, to determine the relative importance of modelling the time of year effect, we performed likelihood ratio tests between the `full' model (shown in Equation~\ref{eqn:monthly_MELR_full}) and a `reduced' model, that incorporates the time of year effect but not its interaction with condition (Equation~\ref{eqn:monthly_MELR_reduced}) and a `null' model that does not consider the time of year effect at all (Equation~\ref{eqn:monthly_MELR_null}). p-values from these tests are reported in Table~\ref{tbl:monthly_LR} with p-values $<$ 0.05 highlighted. 

\begin{equation}
    \begin{split}
   property\_measure \sim & condition \\
    & + sin(toy\_radians) \\
    & + cos(toy\_radians) \\
    & + (1|subject) 
    \end{split}
    \label{eqn:monthly_MELR_reduced}
\end{equation}

\begin{equation}
    \begin{split}
property\_measure \sim & condition \\
    & + (1|subject) 
    \end{split}
    \label{eqn:monthly_MELR_null}
\end{equation}

\begin{table}

    \tiny
    
    \centering
   \begin{tabularx}{\textwidth}{|X|l|l|l|l|l|l|}
    \hline
        \textbf{Period Mean} & Coef. & Std.Err. & z & $P>|z|$ & [0.025 & 0.975] \\ \hline
        Intercept & 24.038 & 0.024 & 989.150 & \textbf{0.000} & 23.990 & 24.085 \\ \hline
        C(Condition)[T.PWE] & 0.016 & 0.026 & 0.611 & 0.541 & -0.036 & 0.068 \\ \hline
        sin\_toy\_radians & 0.027 & 0.018 & 1.518 & 0.129 & -0.008 & 0.061 \\ \hline
        C(Condition)[T.PWE]:sin\_toy\_radians & -0.038 & 0.019 & -2.022 & \textbf{0.043 }& -0.074 & -0.001 \\ \hline
        cos\_toy\_radians & -0.006 & 0.019 & -0.341 & 0.733 & -0.043 & 0.031 \\ \hline
        C(Condition)[T.PWE]:cos\_toy\_radians & 0.014 & 0.020 & 0.715 & 0.475 & -0.025 & 0.053 \\ \hline
        Subject Var & 0.009 & 0.004 & ~ & ~ & ~ & ~ \\ \hline
    \end{tabularx}
    \vspace{5mm}

    \centering
    \begin{tabularx}{\textwidth}{|X|l|l|l|l|l|l|}
    \hline
        \textbf{Period Std} & Coef. & Std.Err. & z & $P>|z|$ & [0.025 & 0.975] \\ \hline
        Intercept & 1.011 & 0.072 & 14.079 & \textbf{0.000} & 0.871 & 1.152 \\ \hline
        C(Condition)[T.PWE] & 0.265 & 0.079 & 3.368 & \textbf{0.001} & 0.111 & 0.419 \\ \hline
        sin\_toy\_radians & -0.036 & 0.024 & -1.531 & 0.126 & -0.083 & 0.010 \\ \hline
        C(Condition)[T.PWE]:sin\_toy\_radians & 0.024 & 0.025 & 0.966 & 0.334 & -0.025 & 0.073 \\ \hline
        cos\_toy\_radians & -0.005 & 0.025 & -0.210 & 0.833 & -0.055 & 0.045 \\ \hline
        C(Condition)[T.PWE]:cos\_toy\_radians & -0.014 & 0.027 & -0.537 & 0.591 & -0.067 & 0.038 \\ \hline
        Subject Var & 0.127 & 0.030 & ~ & ~ & ~ & ~ \\ \hline
    \end{tabularx}
    \vspace{5mm}

    \begin{tabularx}{\textwidth}{|X|l|l|l|l|l|l|}
    \hline
        \textbf{Acrophase Mean} & Coef. & Std.Err. & z & $P>|z|$ & [0.025 & 0.975] \\\hline
        Intercept & 15.394 & 0.405 & 37.978 & \textbf{0.000} & 14.599 & 16.188 \\ \hline
        C(Condition)[T.PWE] & 0.031 & 0.446 & 0.069 & 0.945 & -0.843 & 0.905 \\ \hline
        sin\_toy\_radians & 0.076 & 0.066 & 1.149 & 0.250 & -0.053 & 0.205 \\ \hline
        C(Condition)[T.PWE]:sin\_toy\_radians & -0.104 & 0.069 & -1.504 & 0.132 & -0.240 & 0.032 \\ \hline
        cos\_toy\_radians & 0.084 & 0.070 & 1.194 & 0.232 & -0.054 & 0.222 \\ \hline
        C(Condition)[T.PWE]:cos\_toy\_radians & 0.231 & 0.074 & 3.141 & \textbf{0.002} & 0.087 & 0.376 \\ \hline
        Subject Var & 4.750 & 0.391 & ~ & ~ & ~ & ~ \\ \hline
    \end{tabularx}
    \vspace{5mm}

    \begin{tabularx}{\textwidth}{|X|l|l|l|l|l|l|}
    \hline
        \textbf{Acrophase Std} & Coef. & Std.Err. & z & $P>|z|$ & [0.025 & 0.975] \\ \hline
        Intercept & 0.867 & 0.082 & 10.616 & \textbf{0.000 }& 0.707 & 1.027 \\ \hline
        C(Condition)[T.PWE] & 0.252 & 0.090 & 2.815 & \textbf{0.005} & 0.077 & 0.427 \\ \hline
        sin\_toy\_radians & -0.017 & 0.026 & -0.662 & 0.508 & -0.069 & 0.034 \\ \hline
        C(Condition)[T.PWE]:sin\_toy\_radians & -0.012 & 0.028 & -0.436 & 0.663 & -0.066 & 0.042 \\ \hline
        cos\_toy\_radians & -0.009 & 0.028 & -0.313 & 0.754 & -0.064 & 0.046 \\ \hline
        C(Condition)[T.PWE]:cos\_toy\_radians & -0.026 & 0.029 & -0.873 & 0.383 & -0.084 & 0.032 \\ \hline
        Subject Var & 0.166 & 0.035 & ~ & ~ & ~ & ~ \\ \hline
    \end{tabularx}
    \vspace{5mm}

     \begin{tabularx}{\textwidth}{|X|l|l|l|l|l|l|}
    \hline
        \textbf{Amplitude Mean }& Coef. & Std.Err. & z & $P>|z|$ & [0.025 & 0.975] \\ \hline
        Intercept & 10.951 & 0.592 & 18.511 & \textbf{0.000} & 9.791 & 12.110 \\ \hline
        C(Condition)[T.PWE] & -0.692 & 0.651 & -1.062 & 0.288 & -1.968 & 0.585 \\ \hline
        sin\_toy\_radians & 0.289 & 0.096 & 3.010 & \textbf{0.003} & 0.101 & 0.477 \\ \hline
        C(Condition)[T.PWE]:sin\_toy\_radians & -0.024 & 0.101 & -0.239 & 0.811 & -0.222 & 0.174 \\ \hline
        cos\_toy\_radians & 0.678 & 0.103 & 6.603 & \textbf{0.000} & 0.476 & 0.879 \\ \hline
        C(Condition)[T.PWE]:cos\_toy\_radians & -0.335 & 0.107 & -3.122 & \textbf{0.002} & -0.546 & -0.125 \\ \hline
        Subject Var & 10.119 & 0.541 & ~ & ~ & ~ & ~ \\ \hline
    \end{tabularx}
    \vspace{5mm}

    \begin{tabularx}{\textwidth}{|X|l|l|l|l|l|l|}
   \hline
       \textbf{ Amplitude Std} & Coef. & Std.Err. & z & $P>|z|$ & [0.025 & 0.975] \\ \hline
        Intercept & 2.013 & 0.114 & 17.689 & \textbf{0.000} & 1.790 & 2.236 \\ \hline
        C(Condition)[T.PWE] & -0.027 & 0.125 & -0.216 & 0.829 & -0.271 & 0.217 \\ \hline
        sin\_toy\_radians & 0.010 & 0.039 & 0.268 & 0.788 & -0.066 & 0.086 \\ \hline
        C(Condition)[T.PWE]:sin\_toy\_radians & 0.011 & 0.041 & 0.266 & 0.790 & -0.069 & 0.091 \\ \hline
        cos\_toy\_radians & 0.100 & 0.042 & 2.416 & \textbf{0.016} & 0.019 & 0.182 \\ \hline
        C(Condition)[T.PWE]:cos\_toy\_radians & -0.049 & 0.043 & -1.136 & 0.256 & -0.135 & 0.036 \\ \hline
        Subject Var & 0.316 & 0.046 & ~ & ~ & ~ & ~ \\ \hline
        
    \end{tabularx}
    \caption{Summary of the mixed effects models produced by Equation~\ref{eqn:monthly_MELR_full}.} 
    \label{tbl:monthly_MELR}
\end{table}

\begin{table}[!ht]
    \centering
    \begin{tabular}{|l|l|l|}
     \hline
        ~ & reduced & null \\ \hline
        Period Mean & 0.091 & 0.106 \\ \hline
        Period Std & 0.526 & \textbf{0.028} \\ \hline
        Acrophase Mean & \textbf{0.002} & \textbf{0.0} \\ \hline
        Acrophase Std & 0.634 & \textbf{0.0} \\ \hline
        Amplitude Mean & \textbf{0.008} & \textbf{0.0} \\ \hline
        Amplitude Std & 0.497 & \textbf{0.0} \\ \hline
    \end{tabular}
    \caption{p-values for likelihood ratio tests comparing reduced (Equation~\ref{eqn:monthly_MELR_reduced}) and null (Equation~\ref{eqn:monthly_MELR_null}) models to the full (Equation~\ref{eqn:monthly_MELR_full}) model for each property measure. p $<$ 0.05 reject the null hypothesis and suggests the full model fits the data significantly better and should be used. }
    \label{tbl:monthly_LR}
\end{table}

%RESULTS

For the individual in Figure~\ref{fig:supp_annual_cycles_polar}, period and acrophase averages are consistent over the course of the year, but other properties display seasonal variation.

Looking at the MELR output (Table~\ref{tbl:monthly_MELR}), some seasonal effects were observed for intra-individual average and variability in PWE and controls, most pronounced in intra-individual average amplitude. However, we found no evidence that the reported increase in period and acrophase variability in PWE were driven by seasonal variations. 

In Table~\ref{tbl:monthly_LR}, all rows bar average period have a p-value below 0.05 when compared to the null model, indicating that consideration of seasonal effects helps explain some variance in our circadian properties.  Only average acrophase and amplitude had p-values below 0.05 in the reduced column, implying there may be some interaction between epilepsy and annual variation in intra-individual average acrophase and amplitude.

%DISCUSSION
Intra-individual circadian average and variability values appear to vary subtly over the year in both PWE and controls. However, the relevance of this to epilepsy is not as clear: LR testing identified that average acrophase and amplitude are possibly altered in epilepsy, which could imply that PWE experience more of a seasonal shift in these properties than controls, but this needs to be explored more in future studies with substantially more controls. Most important to this work, we found no evidence of seasonal effects in intra-individual variability of period and acrophase. Thus our main results are unlikely to be driven by seasonal effects.

\end{document}